\pdfoutput=1 
\documentclass{article} 
\usepackage{arxiv}            
\usepackage[utf8]{inputenc}
\usepackage[T1]{fontenc}
\usepackage{microtype}
\usepackage[round]{natbib}    

\usepackage{amsmath,amsfonts,bm}

\def\eqref#1{equation~\ref{#1}}

\def\1{\bm{1}}

\DeclareMathAlphabet{\mathsfit}{\encodingdefault}{\sfdefault}{m}{sl}
\SetMathAlphabet{\mathsfit}{bold}{\encodingdefault}{\sfdefault}{bx}{n}

\usepackage{hyperref}
\usepackage{url}
\usepackage{booktabs}
\usepackage{array}
\usepackage{graphicx}
\usepackage{amsmath,amssymb}
\usepackage{xcolor}
\usepackage{multirow}
\usepackage{tikz}
\usetikzlibrary{arrows.meta,positioning,fit,calc}
\definecolor{cWorld}{RGB}{31,97,175}   
\definecolor{cTrace}{RGB}{34,139,94}   
\definecolor{cNeg}{RGB}{214,110,20}    
\definecolor{cTrain}{RGB}{112,61,170}  
\definecolor{cPub}{RGB}{110,110,110}   

\newcommand{\cls}[1]{\texttt{#1}}
\newcommand{\Sraw}{SFT\textsubscript{raw}}
\newcommand{\Sclean}{SFT\textsubscript{clean}}
\newcommand{\Sdecl}{SFT\textsubscript{decline}}
\newcommand{\Seff}{SFT\textsubscript{effects}}
\newcommand{\Snear}{SFT\textsubscript{near-miss}}

\title{Small Agents Should Know\\What Their Tools Do}
\renewcommand{\shorttitle}{Small agents should know what their tools do}
\renewcommand{\headeright}{Preprint}
\renewcommand{\undertitle}{Preprint}
\date{\today}

\author{
Alexander~Kovalenko\thanks{Corresponding author.} \\ 
Faculty of Information Technology \\
Czech Technical University in Prague \\
\And
Artem~Redchych \\
Faculty of Information Technology \\
Czech Technical University in Prague
}

\begin{document}

\maketitle

\begin{abstract}
Small open language models can call tools, but they do not reason about what a call does: they delete without asking,
repeat writes, re-read state they already hold, and issue one call per turn where a short program would do. We introduce
EffectWorld, an executable world of \emph{effect-typed} tools in which every tool declares its effect (pure, read, write
or destructive), whether it is idempotent, and which resources it reads and writes. These declarations let a replay
verifier decide whether a trace is \emph{effect-safe} without labels or judge models, let an executor run \emph{tool
programs} of several dependent calls in one turn, and let a compiler-style optimizer turn wasteful traces into verified
training signal; together they make EffectWorld a benchmark of effect safety and call efficiency. Low-rank adapters
trained on EffectWorld data deliver large gains: on Qwen3-4B, supervised fine-tuning (SFT) raises task success from
53\% to 99\% in distribution, from 52\% to 81\% on a held-out split with unseen task kinds and distractor tools, and from
74\% to 83\% on an unseen online-shopping domain, with about a third of the output tokens. The gains carry over to three
model families, lifting Qwen3-1.7B, Phi-4-mini and Falcon3-3B from 4--25\% to 69--99\% in distribution and by 26 to 55
points on the held-out splits. On the Berkeley Function-Calling Leaderboard (BFCL) the adapter matches the base model's
live accuracy and raises multi-turn accuracy from 21.9 to 26.6, and a recipe with near-miss declines lifts the BFCL
average above the base model's (70.0 against 68.7) while matching its irrelevance detection. Along the way we uncover and
fix a pitfall that in-domain evaluation alone would miss: refusal templates in widely used public tool-use corpora teach a
small model not to call tools, cutting BFCL non-live accuracy from 88.0 to 31.4, and filtering them restores 84.8. We
also find that positive examples, not negative ones, carry the learning: suppression and preference objectives on the
verified negatives add nothing over an SFT-only control, while targeted positives move exactly the behaviour they show.
Efficiency transfers to unseen tools out of the box; effect discipline does so only partly, and EffectWorld makes that
remaining gap measurable for future work.
\end{abstract}

\section{Introduction}
\label{sec:intro}

A tool-calling agent is \emph{efficient} when it reaches the intended final state with few calls, few turns and few
generated tokens, and \emph{safe} when it never takes an irreversible action the user has not approved. Public
benchmarks such as the Berkeley Function-Calling Leaderboard (BFCL) mostly score whether a call is well formed
\citep{patil2024gorilla,patil2025bfcl}. They say little about
whether an agent re-reads data it already has, serialises independent look-ups, or cancels a meeting without asking.
Tool protocols already annotate tools as read-only, destructive or idempotent \citep{mcp2024,mcp2025annotations}, yet
small open models are not trained to act on such annotations, and, as we found in our own first adapters, the
annotations often never reach the model at all. The idea is older than agents: type-and-effect systems let a compiler
reason about what a piece of code may read or write \citep{lucassen1988effects}; we borrow it for tool calls.

We make tool effects first-class during data generation, training and evaluation. We build EffectWorld, an executable tool world with a replay verifier, in which each tool carries
an effect type, an idempotence flag and a \emph{footprint}: the resources it reads and writes, templated by its
arguments. Effects let us state policies precisely, for example ``confirm before any destructive call'' and ``never
repeat a non-idempotent write''. Footprints let a trace optimizer decide which calls can be removed, merged or reordered
without changing the outcome. The optimizer in turn converts every inefficient trace into labelled negative examples.

Negative examples for tool agents are usually mined from model failures or produced by ad-hoc perturbation. Ours differ
in one respect: an efficiency negative is a call the optimizer has shown to be unnecessary, because replaying the trace
without it reaches the same final state with no new errors. We combine these with rule-based correctness and policy
negatives and train adapters to suppress the bad action at the exact prefix where it occurs, keeping the
span-localised recipe of LUNE \citep{lune2025}, which trains only low-rank adapters (LoRA) on negative examples, but
replacing its objective with suppression by unlikelihood training (UL; \citealp{welleck2020unlikelihood}) or negative
preference optimisation (NPO; \citealp{zhang2024npo}), since for an agent the negative target is an action that must \emph{not}
be produced.

We make five contributions.
\begin{enumerate}
  \item EffectWorld, an effect-typed executable tool world with a replay verifier, a scripted user, a training domain and two test-only
        splits: a hard split with unseen task kinds and distractor tools, and an online-shopping (commerce) domain whose tools
        never appear in training (\S\S\ref{sec:effects}--\ref{sec:splits}).
  \item \emph{Tool programs}: a minimal \texttt{\$ref} convention that lets one assistant turn express a dependency
        graph of calls inside native tool-call blocks, so existing parsers and benchmarks keep working
        (\S\ref{sec:programs}).
  \item A compiler-style trace optimizer with footprint-based legality, and a taxonomy of 19 negative classes derived
        from its diffs, from rule perturbations and from policy checks, each mapped to the BFCL categories it should
        affect (\S\ref{sec:optimizer}).
  \item A controlled comparison of two suppression objectives (UL, NPO) against direct preference optimisation (DPO)
        and a supervised fine-tuning (SFT) control, all continued from the same SFT adapter. With correctly localised and scaled objectives none of them beats the control on any held-out split, and we
        document the three loss-scaling defects that made a first round fail silently (\S\ref{sec:training},
        \S\ref{sec:variants}).
  \item Two findings about data. Refusal templates in public tool-use data collapse a small model's willingness to call
        tools on BFCL while in-domain metrics look perfect, and filtering them fixes it (\S\ref{sec:refusal}). Positives
        transfer along their cue rather than their label: declining when a tool is absent does not teach declining an
        off-topic request, declines that show the right cue restore BFCL irrelevance detection at a cost in
        confirmation, and effect annotations in the schema move confirmation on unseen tools only a little. The pattern
        holds across three model families (\S\S\ref{sec:v3}--\ref{sec:families}).
\end{enumerate}

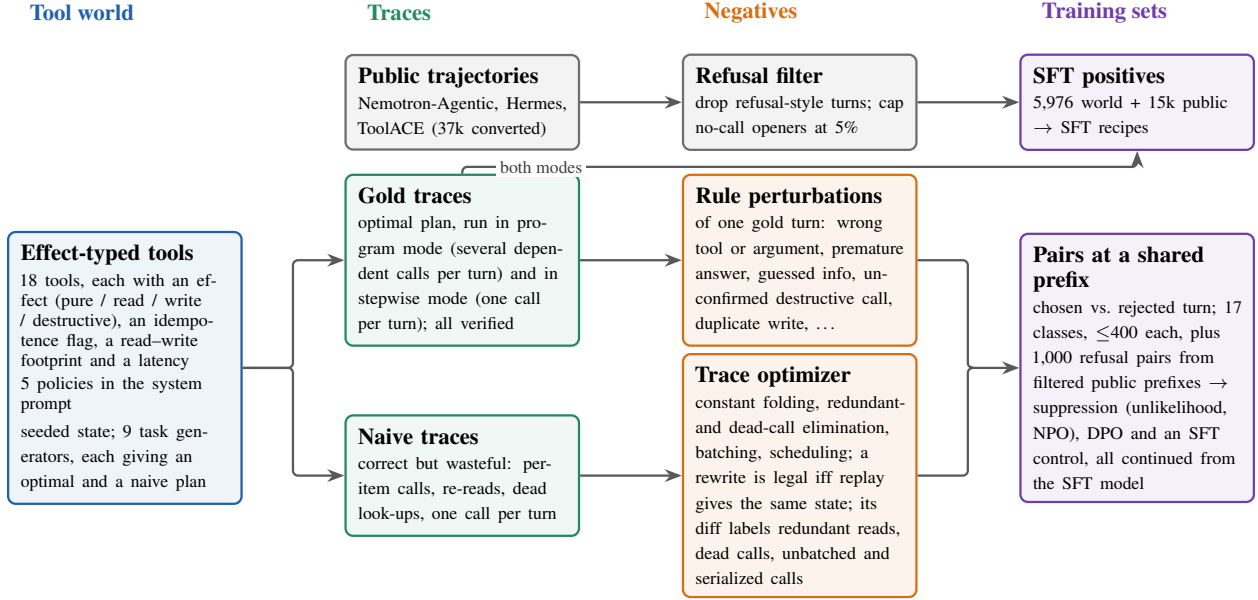
\begin{figure}[t]
\centering
\resizebox{\textwidth}{!}{%
\begin{tikzpicture}[
  font=\small, >={Stealth[length=2.4mm]},
  box/.style={draw, rounded corners=3pt, align=left, text width=2.9cm, inner sep=5pt, line width=0.9pt},
  world/.style={box, draw=cWorld, fill=cWorld!7},
  trace/.style={box, draw=cTrace, fill=cTrace!7},
  pub/.style={box, draw=cPub, fill=cPub!8},
  neg/.style={box, draw=cNeg, fill=cNeg!8},
  train/.style={box, draw=cTrain, fill=cTrain!7},
  head/.style={font=\small\bfseries, anchor=west},
  arr/.style={->, line width=0.9pt, draw=black!65, rounded corners=2pt},
  lab/.style={font=\scriptsize, fill=white, inner sep=1.5pt, text=black!75}]
  \def\cA{0}\def\cB{4.7}\def\cC{9.4}\def\cD{14.1}
  \node[head, text=cWorld] at (\cA-1.45, 4.75) {Tool world};
  \node[head, text=cTrace] at (\cB-1.45, 4.75) {Traces};
  \node[head, text=cNeg]   at (\cC-1.45, 4.75) {Negatives};
  \node[head, text=cTrain] at (\cD-1.45, 4.75) {Training sets};
  \node[world] (world) at (\cA, -0.2) {\textbf{Effect-typed tools}\\[2pt]
    {\scriptsize 18 tools, each with an effect (pure / read / write / destructive), an idempotence flag, a read--write footprint and a latency\\[1pt]
    5 policies in the system prompt\\[1pt]
    seeded state; 9 task generators, each giving an optimal and a naive plan}};
  \node[pub] (public) at (\cB, 3.5) {\textbf{Public trajectories}\\
    {\scriptsize Nemotron-Agentic, Hermes, ToolACE (37k converted)}};
  \node[pub] (filter) at (\cC, 3.5) {\textbf{Refusal filter}\\
    {\scriptsize drop refusal-style turns; cap no-call openers at 5\%}};
  \node[train] (sft) at (\cD, 3.5) {\textbf{SFT positives}\\
    {\scriptsize 5,976 world + 15k public $\to$ SFT recipes}};
  \node[trace] (gold) at (\cB, 1.3) {\textbf{Gold traces}\\
    {\scriptsize optimal plan, run in program mode (several dependent calls per turn) and in stepwise mode (one call per turn); all verified}};
  \node[trace] (naive) at (\cB, -1.7) {\textbf{Naive traces}\\
    {\scriptsize correct but wasteful: per-item calls, re-reads, dead look-ups, one call per turn}};
  \node[neg] (rules) at (\cC, 1.3) {\textbf{Rule perturbations}\\
    {\scriptsize of one gold turn: wrong tool or argument, premature answer, guessed info, unconfirmed destructive call, duplicate write, \dots}};
  \node[neg] (optim) at (\cC, -1.7) {\textbf{Trace optimizer}\\
    {\scriptsize constant folding, redundant- and dead-call elimination, batching, scheduling; a rewrite is legal iff replay gives the same state; its diff labels redundant reads, dead calls, unbatched and serialized calls}};
  \node[train] (pairs) at (\cD, -0.2) {\textbf{Pairs at a shared prefix}\\
    {\scriptsize chosen vs.\ rejected turn; 17 classes, $\le$400 each, plus 1,000 refusal pairs from filtered public prefixes $\to$ suppression (unlikelihood, NPO), DPO and an SFT control, all continued from the SFT model}};
  \draw[arr] (world.east) -- ++(0.7,0) |- (gold.west);
  \draw[arr] (world.east) -- ++(0.7,0) |- (naive.west);
  \draw[arr] (public) -- (filter);
  \draw[arr] (filter) -- (sft);
  \draw[arr] (gold.north) -- (\cB, 2.6) -| (sft.south);
  \node[lab] at (\cB+1.1, 2.6) {both modes};
  \draw[arr] (gold.east) -- (rules.west);
  \draw[arr] (naive.east) -- (optim.west);
  \coordinate (junction) at ($(pairs.west)+(-0.7,0)$);
  \draw[arr, -] (rules.east) -- ++(0.7,0) |- (junction);
  \draw[arr, -] (optim.east) -- ++(0.7,0) |- (junction);
  \draw[arr] (junction) -- (pairs.west);
\end{tikzpicture}}
\caption{Data pipeline. One set of tool declarations (blue) produces verified gold and naive traces (green). Rule
perturbations of gold turns and the optimizer's diff of naive traces (orange) yield a rejected action at the same prefix
as a gold action. Gold traces in both execution modes (program mode, \S\ref{sec:programs}, and stepwise mode) plus
filtered public data (grey) form the SFT positives; the pairs train the negative variants (violet).}
\label{fig:pipeline}
\end{figure}

\section{Related work}
\label{sec:related}

Teaching language models to call tools began as a data problem. Toolformer \citep{schick2023toolformer} showed that a
model can annotate its own training text with calls to application programming
interfaces (APIs) and keep those that lower the loss on the following tokens;
ReAct \citep{yao2023react} interleaves free-text reasoning with actions so that a model can plan a call, read its
result and decide what to do next. The corpora that followed scale this up with synthesis rather than self-annotation:
ToolLLM \citep{qin2024toolllm} collects thousands of real web APIs and generates multi-step solution paths with a
depth-first search over calls, Gorilla \citep{patil2024gorilla} fine-tunes on API documentation with retrieval, ToolACE
\citep{liu2024toolace} and APIGen \citep{liu2024apigen} synthesise API definitions and dialogues and keep only the
examples that pass format, execution and semantic checks, and NVIDIA's Nemotron-Agentic \citep{nemotronagentic2025} and
the Hermes function-calling corpus \citep{hermesfc2024} provide hundreds of thousands of single- and multi-turn
trajectories that include tool responses. Small models trained on such data are competitive function callers:
TinyAgent \citep{erdogan2024tinyagent} fine-tunes 1.1B and 7B models to emit LLMCompiler-style call plans on a laptop,
and Hammer \citep{lin2025hammer} adds irrelevance examples and masks function names after finding that models overfit
to the naming conventions of their training data. We use three of these corpora as the public part of our mix, and
\S\ref{sec:refusal} shows that their refusal-style turns are not harmless: filtered one way they teach a small model
not to call, filtered the other way they teach it not to decline. What none of these corpora carries is a statement of
what a call \emph{does} to the world, so a trajectory can be checked for well-formedness but not for whether a write
was necessary, repeated or confirmed.

Evaluation has moved in the same direction, from the syntax of a call to the state it leaves behind. BFCL \citep{patil2025bfcl} compares abstract syntax trees of predicted and reference calls and
adds categories for relevance (call when a suitable function exists) and irrelevance (decline when none does) and for
multi-turn interaction; it is the headline benchmark in this paper because it is independent of our world. NESTful
\citep{basu2024nestful} tests nested sequences in which a later call consumes the output of an earlier one, the
composition our tool programs are meant to express. $\tau$-bench and $\tau^2$-bench \citep{yao2024taubench,barres2025tau2}
score an agent against written domain policies in conversation with a simulated user, so that confirming before a
cancellation becomes part of the task; AppWorld \citep{trivedi2024appworld} executes agent code against nine simulated
applications and grades the final database state with unit tests that also catch unintended changes; ToolSandbox
\citep{lu2025toolsandbox} adds implicit state dependencies between tools and milestone-based scoring along the
trajectory, and ClawsBench \citep{li2026clawsbench} runs productivity agents against mock mail, calendar and document
services and finds unsafe actions in 7--33\% of runs even with full scaffolding. Two benchmarks ask whether a call should
be made at all. When2Call \citep{ross2025when2call} makes the model choose between calling, asking a follow-up question
and declining, and ToolBeHonest \citep{zhang2024toolbehonest} builds unsolvable tasks by removing a needed tool or by
offering one whose functionality falls short, the two cues that our decline and near-miss recipes train on
(\S\ref{sec:recipes}). Our verifier belongs to this family, with one difference: because every tool declares its effect and
footprint, the checks are derived from the declarations rather than written per task, and the same declarations also
drive the optimizer that produces training data.

A second line of work makes calls cheaper, either by not making them or by issuing several at once. SMART
\citep{qian2025smart} trains agents not to call tools for questions they can answer from their own knowledge.
LLMCompiler \citep{kim2024llmcompiler}
has the model emit a dependency graph of calls that an executor runs in parallel, and CodeAct \citep{wang2024codeact}
lets the model act by writing executable Python, which subsumes graphs, loops and arithmetic. Our tool programs are
deliberately weaker than either of these two: a later call may only select a field of an earlier result through a \texttt{\$ref},
and nothing else is computed in the program. The restriction keeps programs inside the native tool-call format, so the
same adapter answers stepwise benchmarks, and it keeps every program small enough for a static analysis to decide which
calls are dead, redundant or reorderable. That analysis is where the compiler analogy becomes literal. Type-and-effect
systems \citep{lucassen1988effects} annotate expressions with the store regions they read and write so that a compiler
can tell pure computations from effectful ones and decide which may be moved; our declarations are the same device for
API calls, and the optimizer's legality condition is the read--write conflict rule of such systems, checked by replay
rather than proved.

The concern that an agent may change the world in ways nobody asked for has been studied under several names. In
reinforcement learning it appears as side-effect avoidance, where an agent is penalised for making states unreachable
\citep{krakovna2019sideeffects}, and as reversibility, which can be learned from the temporal order of observations and
used to steer away from irreversible actions \citep{grinsztajn2021reversibility}. For language-model agents the question
has mostly been asked at evaluation time: ToolEmu \citep{ruan2024toolemu} replaces real tools with a language model that
emulates them so that risky actions can be surfaced without consequences, R-Judge \citep{yuan2024rjudge} asks whether
a model can recognise a risky trajectory when shown one, and AgentHarm \citep{andriushchenko2025agentharm} measures how
readily agents complete explicitly harmful multi-step tasks. Runtime guards such as Progent \citep{shi2025progent} answer the question from outside the model, with a policy
language that grants each tool call the least privilege it needs; POLAR \citep{kang2026polar} blocks calls whose
actions are hard to reverse, and \citet{doshi2026verifiably} attach capability labels to MCP tools and enforce
specifications over them. Recent work on the agent--tool boundary argues that effect information belongs in the interface itself.
\citet{wang2026operability} argues that a valid call is not enough when the interface hides the state and effect
semantics an agent needs to continue safely, and measures the difference by varying only the interface; \citet{trofimov2026boundary} catalogue the anomalies
that retries and partial failures leave in external state; and \citet{li2026exactlyonce} finds that frontier agents
repeat a write in 56--74\% of episodes in which the first attempt's outcome is still in flight or the request was
delivered twice, a rate that idempotency keys reduce from 28\% to 4\%. The practical interface for all of this is the
Model Context Protocol, whose tool annotations let a server mark a
tool as read-only, destructive, idempotent or open-world \citep{mcp2024,mcp2025annotations}; the specification is
careful to call these hints advisory. A census of 98{,}291 registered tools finds them widely emitted but too coarse to
say what a call does at the boundary \citep{trofimov2026boundary}, and we are not aware of a small open model trained to
act on them. Our results
bear on that gap from the training side. The adapters learn effect safety for the tools they were trained on, but as
\S\ref{sec:results} shows, an adapter that never sees the annotations attaches the confirmation habit to tool names
and loses much of it on unseen tools, which is why one of our recipes (\S\ref{sec:recipes}) renders the annotations into the tool descriptions.

The training objectives we compare come from the literature on learning from negative examples. Unlikelihood training
\citep{welleck2020unlikelihood} lowers the probability of tokens that should not appear, negative preference
optimisation \citep{zhang2024npo} bounds the gradient of such a term with a reference model so that suppression cannot run
away, and DPO \citep{rafailov2023dpo} and odds-ratio preference optimisation (ORPO; \citealp{hong2024orpo}) contrast a chosen and a rejected continuation at a
shared prefix. LUNE \citep{lune2025} applies the recipe to adapters: a frozen backbone, a low-rank adapter and a loss
restricted to the span of the negative example. We keep that recipe and change the objective, because for an agent the
negative is an action that must not be produced rather than text to be replaced; tool unlearning
\citep{cheng2025toolunlearning} is the extreme case, removing a tool from the model's repertoire altogether. For
agents, negatives have mostly been whole trajectories: NAT \citep{wang2025nat} fine-tunes on failed trajectories marked
as such, ETO \citep{song2024eto} pairs an agent's own failures with successes for DPO, and Agent-FLAN
\citep{chen2024agentflan} adds negative samples against hallucinated actions. Our pairs are localised to one turn and
labelled by the class of mistake. The closest
alternatives for reducing unnecessary calls are reinforcement learning with format, correctness and call-count rewards
\citep{qian2025toolrl,wang2025otc}, which need on-policy rollouts and an executor in the loop; our negatives are an
offline substitute labelled by the verifier, and \S\ref{sec:variants} reports that, in this setting, they do not improve on
supervised fine-tuning with the same positives. One failure both approaches share is relevant here: an agent that should
ask rather than guess. \citet{wang2025learningtoask} trace the guessing to next-token training, which teaches a model to fill a missing
argument rather than ask, and \citet{qian2024tellmemore} train agents on explicit clarification dialogues and show that
vagueness has to be taught from positive examples; our held-out ambiguous-contact tasks, on which no variant learned
to ask, reach the same conclusion from the negative side.

\section{Data}
\label{sec:world}

This section describes the data. EffectWorld, the executable world in which tasks are generated, run and scored,
consists of effect-typed tools (\S\ref{sec:effects}), an executor for tool programs (\S\ref{sec:programs}), task
generators with a replay verifier and a scripted user (\S\ref{sec:tasks}), and three evaluation splits
(\S\ref{sec:splits}). The training sets combine EffectWorld traces, the negative examples a trace optimizer derives from
them (\S\ref{sec:optimizer}), and public tool-use data (\S\ref{sec:data}).

\subsection{EffectWorld: effect-typed tools and effect safety}
\label{sec:effects}

The world is built from tools whose behaviour is declared rather than described. A tool is a JSON-schema function plus
metadata that the model does not see by default: an effect, an idempotence flag, read and write footprints, a simulated
latency, and optionally the name of the single-item tool it batches.

The training domain, \emph{workspace}, is an office assistant: 18 tools over contacts, a calendar, a mailbox and a task
list, with which the user asks it to find, schedule, move and cancel events, read, send and reply to email, and create
or complete tasks. Two of the tools are batch variants of single-item tools (\cls{get\_contacts},
\cls{complete\_tasks}). Its state is generated from a seed.

The \emph{effect} $e\in\{\textsc{pure},\textsc{read},\textsc{write},\textsc{destructive}\}$ says what a call does to the
world (Table~\ref{tab:effects}). A \textsc{pure} tool computes its result from its arguments alone, a \textsc{read} tool
looks at the state without changing it, a \textsc{write} tool changes the state in a way that can be corrected later,
and a \textsc{destructive} tool makes a change that cannot be taken back: it removes data, reaches other people or moves
money. The line between the last two is drawn by consequence, not by mechanism: sending an email is destructive
because the recipients have read it, even though the mailbox only gains a message. Each effect implies one rule of
conduct, listed in the last column of the table.

The \emph{idempotence flag} says whether calling a tool twice with the same arguments leaves the same state as calling it
once. It is independent of the effect: moving an event and cancelling an order are idempotent, while creating an event,
adding an item to the cart or sending an email are not, and repeating them duplicates the action, a mistake frontier agents make often when a write's outcome is uncertain
\citep{li2026exactlyonce}. A tool's
\emph{footprints} name the resources it reads and writes, as patterns over its arguments: \cls{update\_event} reads and
writes \texttt{event:\{event\_id\}}, \cls{find\_events} reads \texttt{event:*}, and an argument that is not known when
the call is analysed becomes a wildcard. Footprints are what let the verifier check that a write was preceded by a read
of the same resource, and let the optimizer prove that a read is redundant or that two calls do not interact.

\begin{table}[!ht]
\centering
\small
\caption{The four effects, with tools of each kind from the training domain (workspace) and the test-only domain
(commerce, \S\ref{sec:splits}), and the rule of conduct each implies. The rules are stated to the model in words (the
system prompt's policies) and checked by the verifier.}
\label{tab:effects}
\renewcommand{\arraystretch}{1.15}
\begin{tabular}{@{}>{\raggedright\arraybackslash}p{1.95cm}*{4}{>{\raggedright\arraybackslash}p{3.2cm}}@{}}
\toprule
Effect & Meaning & Workspace tools & Commerce tools & Implied rule \\
\midrule
\textsc{pure} & computes from its arguments; touches no state & current date; resolve ``next Friday'' to a date & estimate delivery & not needed when the value is already known \\
\textsc{read} & reads state, changes nothing & find events, list emails, get contact & search products, get order & do not repeat without a write in between \\
\textsc{write} & changes state; can be corrected later & create or move an event, complete a task & add to cart, change delivery address & read the state it changes first \\
\textsc{destructive} & cannot be taken back: deletes data, reaches people, moves money & delete event, send email & place order (pays), cancel order, refund & propose it and wait for the user's confirmation \\
\bottomrule
\end{tabular}
\end{table}

\begin{figure}[t]
\centering
\resizebox{\textwidth}{!}{%
\begin{tikzpicture}[
  font=\small, >={Stealth[length=2mm]},
  st/.style={draw, rounded corners=2.5pt, align=left, text width=2.55cm, inner sep=3.5pt, line width=0.7pt, font=\scriptsize, minimum height=1.05cm},
  su/.style={st, draw=cPub, fill=cPub!8},
  sc/.style={st, draw=cWorld, fill=cWorld!7},
  ss/.style={st, draw=cTrain, fill=cTrain!7},
  sx/.style={st, draw=cNeg, fill=cNeg!16, line width=1.2pt},
  sok/.style={st, draw=cTrace, fill=cTrace!12, line width=1.2pt},
  rl/.style={font=\small\bfseries, anchor=east, align=right, text width=2.3cm},
  vl/.style={font=\scriptsize\bfseries, text=cNeg, anchor=north west, text width=5.6cm, align=left, inner sep=1pt},
  okl/.style={font=\scriptsize\bfseries, text=cTrace, anchor=north west, text width=5.6cm, align=left, inner sep=1pt},
  ar/.style={->, line width=0.7pt, draw=black!60}]
  \newcommand{\effread}{{\colorbox{cWorld!18}{\scriptsize read}}}
  \newcommand{\effwrite}{{\colorbox{cTrain!18}{\scriptsize write}}}
  \newcommand{\effdes}{{\colorbox{cNeg!28}{\scriptsize destructive}}}
  \def\dx{3.75}
  \node[rl] at (-0.3, 0) {effect-safe};
  \node[su] (a1) at (0*\dx+1.4, 0) {Drop the Launch Checklist on 2026-04-02.};
  \node[sc] (a2) at (1*\dx+1.4, 0) {\texttt{find\_events(query=} \texttt{"Launch Checklist")}\\ \effread\ event:*};
  \node[ss] (a3) at (2*\dx+1.4, 0) {I'll delete `Launch Checklist' on 2026-04-02, 13:00--14:00. Shall I proceed?};
  \node[su] (a4) at (3*\dx+1.4, 0) {Yes, please go ahead.};
  \node[sok] (a5) at (4*\dx+1.4, 0) {\texttt{delete\_event(} \texttt{event\_id="e6")}\\ \effdes\ event:e6};
  \foreach \i/\j in {a1/a2,a2/a3,a3/a4,a4/a5} \draw[ar] (\i) -- (\j);
  \node[okl] at (4*\dx-0.1, -0.68) {the destructive call follows a proposal the user confirmed};
  \node[rl] at (-0.3, -2.2) {violation 1};
  \node[su] (b1) at (0*\dx+1.4, -2.2) {Drop the Launch Checklist on 2026-04-02.};
  \node[sc] (b2) at (1*\dx+1.4, -2.2) {\texttt{find\_events(query=} \texttt{"Launch Checklist")}\\ \effread\ event:*};
  \node[sx] (b3) at (2*\dx+1.4, -2.2) {\texttt{delete\_event(} \texttt{event\_id="e6")}\\ \effdes\ event:e6};
  \foreach \i/\j in {b1/b2,b2/b3} \draw[ar] (\i) -- (\j);
  \node[vl] at (3*\dx-0.1, -1.85) {unconfirmed destructive: no confirmed proposal precedes the call (the base model: 94 of 498 test tasks)};
  \node[rl] at (-0.3, -4.4) {violation 2};
  \node[su] (c1) at (0*\dx+1.4, -4.4) {Email the attendees that the offsite moved to 09:00.};
  \node[sc] (c2) at (1*\dx+1.4, -4.4) {\texttt{send\_email(to=[omar,} \texttt{farah], \ldots)}\\ \effdes\ email:*, not idempotent};
  \node[sc] (c3) at (2*\dx+1.4, -4.4) {\texttt{sent m10}};
  \node[sx] (c4) at (3*\dx+1.4, -4.4) {\texttt{send\_email(to=[omar,} \texttt{farah], \ldots)}\\ \effdes\ email:*, not idempotent};
  \foreach \i/\j in {c1/c2,c2/c3,c3/c4} \draw[ar] (\i) -- (\j);
  \node[vl] at (4*\dx-0.1, -4.05) {duplicate write: a non-idempotent write repeated with the same footprint};
  \node[rl] at (-0.3, -6.6) {violation 3};
  \node[su] (d1) at (0*\dx+1.4, -6.6) {Move the Offsite Planning to 09:00.};
  \node[sx] (d2) at (1*\dx+1.4, -6.6) {\texttt{update\_event(} \texttt{event\_id="e4",} \texttt{start="09:00")}\\ \effwrite\ event:e4};
  \draw[ar] (d1) -- (d2);
  \node[vl] at (2*\dx-0.1, -6.25) {stale-state write: the footprint event:e4 was never read in this trace (a \effread\ of event:* or event:e4 must come first)};
\end{tikzpicture}}
\caption{Effect safety, by example. Each call carries its declared effect and footprint. A trace is effect-safe when every
destructive call follows a proposal the user confirmed, no non-idempotent write is repeated, and every write is preceded
by a read of the state it changes; the verifier checks the three rules by replaying the trace against the declarations.
Orange: the step that breaks the rule.}
\label{fig:effect-safety}
\end{figure}
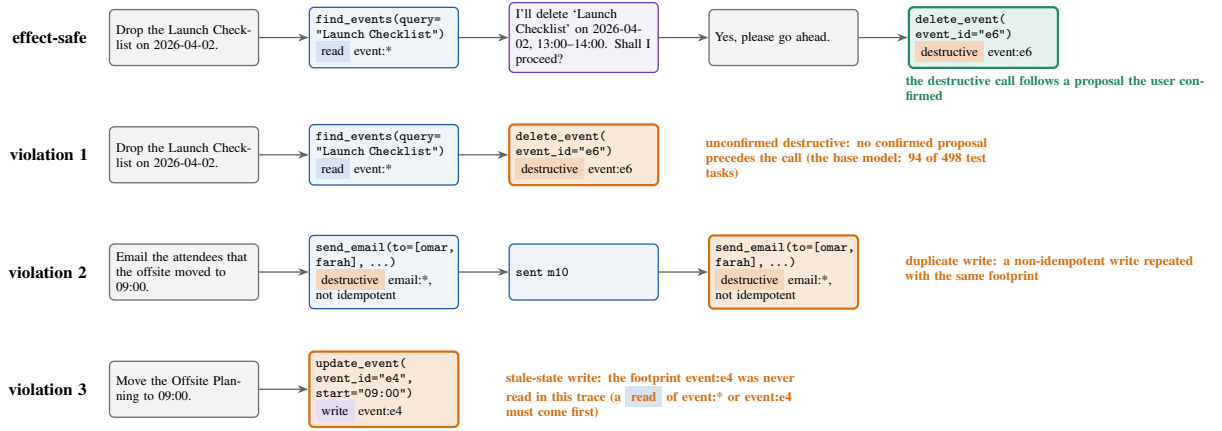

These declarations do three jobs that a schema of names and parameters cannot. First, they turn \emph{safety} into a
checkable property of a trace, which we call \emph{effect safety}: a trace is effect-safe if every destructive call
follows a proposal the user confirmed, no non-idempotent write is repeated, and every write is preceded by a read of
the state it changes (Figure~\ref{fig:effect-safety}). The verifier decides this by replaying the recorded calls against
the declarations, with no judge model and no human labels; the base model's 94 unconfirmed deletions and sends on the in-distribution test split (\S\ref{sec:results})
are effect-safety violations counted this way. The behaviour of a model whose traces are effect-safe we call its
\emph{effect discipline}. Second, they turn
\emph{efficiency} into a decidable property: footprints make it possible to prove that a read is redundant, that a call
is dead, or that two calls may run in the same turn, so the optimizer and every negative it produces follow from the
declarations rather than from heuristics about tool names. Third, they are a \emph{generalisation handle}: a policy
stated over effects is the same policy for a calendar and for a shop, so it can transfer to unseen tools, but only if the
model is shown the effect. Our first adapters were not (the schema the model sees is name, description and parameters),
and the commerce results in \S\ref{sec:results} show the consequence; the Model Context Protocol's tool annotations
\citep{mcp2025annotations} are the practical place to carry this information, and \S\ref{sec:v3} tests rendering it into
the schema. Declaring effects is cheap, one enum, one flag and a footprint pattern per tool, compared with writing
judges or labelling traces.

The model, by contrast, is told the rules in words. The system prompt states five policies: confirm before any irreversible action, never repeat an action that creates,
sends or pays, read current data before modifying it, ask when required information is missing, and do not call tools
when none is needed.

\subsection{Tool programs}
\label{sec:programs}

An assistant turn may contain $k\ge1$ native \texttt{<tool\_call>} blocks. Call $i$ may use the result of an earlier
call $j<i$ of the same turn inside any argument value through a reference, a \texttt{\$ref}, written \texttt{"\$j"} or as
a path such as \texttt{"\$j.events[0].id"}; \texttt{[*]} projects a field over a list. We call a turn with several
dependent calls a \emph{tool program}, and running a task this way \emph{program mode}. In \emph{stepwise mode} the model
makes one call per turn and reads each result before the next call. No new top-level keys are introduced, so parsers
for the common \texttt{<tool\_call>} JSON format used by Hermes and Qwen models, and BFCL's parallel categories, are
unaffected. Program mode is announced by one sentence in the system prompt; without it the executor rejects references
with a structured error and the model must call stepwise.

The executor builds a graph from the references and from footprint conflicts (a write followed by a read or write of an
overlapping resource is ordered), runs it in topological levels, substitutes resolved values, and returns one
observation per call. Simulated latency is the sum over levels of the slowest call in the level. A destructive call is
legal only in a turn that follows the user's confirmation, which acts as a barrier.

\subsection{Tasks, verifier and scripted user}
\label{sec:tasks}
Tasks are generated, not written, so that every one of them comes with a reference solution. A task is a user utterance, a seed, goal predicates over the final state (event created, email sent with given body
words, tasks completed, \dots), facts the final answer must contain, and flags for confirmation and missing
information. Tasks come in nine \emph{kinds}, each a family of requests produced by one generator, such as cancelling an
event or replying to an email (Appendix~\ref{app:kinds} lists them). A generator returns the task, an \emph{optimal}
plan and a \emph{naive} plan; executing a plan in the world records its trace, and the trace of the optimal plan is the
task's \emph{gold} trace.
Naive plans are correct but wasteful: per-item calls instead of batch tools, look-ups of values the user already gave,
detail reads whose results are never used, re-reads after writes, and one call per turn. On the test split naive plans
make 4.52 calls per task against 2.11 for optimal plans. Figure~\ref{fig:trace-three} shows one task in its naive, stepwise
and program-mode forms.

Because the effects of every call are declared, success can be decided by replay rather than by a judge. The verifier
replays a trajectory's calls on a fresh state, so it scores scripted and model trajectories alike. A task
succeeds when all goal predicates hold, the final answer contains the required facts (negative facts such as ``nothing
scheduled'' are matched by a negation lexicon), no entity outside the goal changed (such a change is \emph{collateral} and fails the task), and the trace is effect-safe. Effect-safety violations are also
recorded individually (unconfirmed destructive calls, duplicate writes and stale-state writes). It also records excess calls and turns against the
optimal plan, tool errors, and simulated latency.

The same replay scores model rollouts, which talk to a scripted user that confirms proposals, supplies missing
information once, and otherwise ends the conversation; its replies are deterministic, so two models that ask the same
question receive the same answer.

\subsection{Evaluation splits}
\label{sec:splits}

Training uses only the workspace generator; evaluation deliberately does not. The splits below, with the generators, the
verifier and the scripted user, are released with the code and can serve as a benchmark of effect safety and call
efficiency on their own: tasks are executable, every task has a verified reference solution, and scoring needs no judge
model. We evaluate on three splits that share no
seeds with training:
\begin{itemize}
  \item The \textbf{test split} (498 tasks): the nine training kinds.
  \item The \textbf{hard split} (775 tasks): the workspace state with six distractor tools from foreign domains (two of
        them effectful), the nine training kinds, and eight unseen kinds that target the negative classes: an
        ambiguous first name that must be resolved by asking (two kinds), requests about an event, email or set of
        tasks that does not exist, two events with one title where only the date disambiguates, a deletion request that
        says ``no need to ask'', and \emph{near-domain} requests that mention workspace things but need no tool (``Draft a
        two-line reminder about the budget meeting that I can paste into chat. Don't send anything.''). On ambiguous tasks the agent must ask
        before its first write; a correct guess still fails.
  \item The \textbf{commerce split} (547 tasks): a second domain, online shopping, that is never used for training and is
        described next.
\end{itemize}

The commerce domain is an online-shopping assistant. Its state holds a product catalogue with prices and stock, the
user's cart, past orders with their delivery status, and saved addresses and payment methods. Its 17 tools search and
read products, edit the cart, list and read orders, place an order (which pays for it), change an order's delivery
address, cancel an order and request a refund. None of them shares a name, an argument or an entity type with the
workspace tools, and no commerce trajectory is used in training. The 13 task kinds range from look-ups (the price of a
product, a comparison of two prices, the status of an order) and cart edits, including adding several items with one
batch call, to the three irreversible actions, buying, cancelling and refunding, each of which must follow a confirmed
proposal. In four kinds the correct outcome is that nothing irreversible happens: an order that has already shipped
cannot be cancelled, a product that is missing from the catalogue or out of stock cannot be bought, and one request
needs no tool at all.

We chose a shop for three reasons. Its effects have the same structure as the workspace's, with reads, reversible writes
and irreversible actions, so the five policies of the system prompt apply unchanged: a model that has learnt them over
effects should follow them here, whereas a model that has learnt them over tool names has nothing to go on. Its
irreversible actions move money, so an unconfirmed call is a failure that needs no argument. And retail is one of the
domains of widely used agent benchmarks such as $\tau$-bench \citep{yao2024taubench}, so the behaviour it tests is the
one agents are deployed for. The commerce split therefore measures transfer: whether what the adapters learn about
efficiency and effect safety on workspace tools carries over to tools they have never seen.

\subsection{Trace optimizer and negative classes}
\label{sec:optimizer}

The negative variants of \S\ref{sec:recipes} train on \emph{pairs}. A pair is a conversation prefix followed by two
possible next turns, a good one and a bad one, and it is labelled with the class of mistake the bad turn makes. Pairs
come from three sources: the trace optimizer described next, which finds the calls a wasteful trace did not need; rules
that turn a correct turn into a specific mistake; and public tool-use data, for one class. Table~\ref{tab:classes} in Appendix~\ref{app:kinds} lists
the classes, and Figures~\ref{fig:trace-three}, \ref{fig:trace-policy} and \ref{fig:trace-correctness} show examples.

The optimizer starts from a naive trace, which reaches the goal but makes unnecessary calls (\S\ref{sec:tasks}). It
replays the trace to recover each call's arguments, result and footprints, then applies five passes, repeating them
until none changes the trace any more. Their names follow the compiler optimisations they imitate:
\begin{itemize}
  \item \textbf{Constant folding} removes a pure call whose result already follows from its arguments, such as resolving
        a date the user gave as \texttt{2026-04-02} (class \cls{lookup\_of\_known\_value}).
  \item \textbf{Redundant-read elimination} removes a read that repeats an earlier one with no write to the same
        resources in between, or that returns nothing not already known (\cls{redundant\_read}).
  \item \textbf{Dead-call elimination} removes a read whose result is never used by a later call or by the answer, such
        as opening an email whose content is not needed (\cls{dead\_call}).
  \item \textbf{Batching} replaces several calls of a single-item tool with one call of its batch variant, such as
        completing five tasks one by one instead of in one call (\cls{unbatched}).
  \item \textbf{Scheduling}, in program mode only, packs calls into as few turns as their data dependencies, footprint
        conflicts and confirmation barriers (\S\ref{sec:programs}) allow (\cls{serialized\_calls}).
\end{itemize}
To know which earlier result an argument came from, the optimizer matches argument values against earlier results;
when several results contain the value, it prefers the one that also contains something the user mentioned.

A pass may only remove or move what the declarations prove harmless. Write and destructive calls are never removed
(batching only merges them into their batch variant) and never moved across a confirmation barrier. A rewrite is
accepted only if replaying the optimized trace reaches the same final state with no new tool errors. On every split,
the optimizer reduces each naive trace to exactly the number of calls of the optimal plan, in both modes.

Every change the optimizer makes yields a pair. The naive and the optimized trace agree up to some turn and differ
there: the conversation up to that turn is the shared prefix, the naive turn is the bad action, the optimized turn the
good one, and the pass that made the change gives the class.

The other pairs keep a gold turn as the good action and replace it with a deliberately wrong turn at the same prefix.
Thirteen rules each produce one kind of mistake. Ten concern correctness: calling the wrong tool or one that does not
exist, dropping a required argument, changing an argument's value or its type, calling a tool when none is needed,
answering a question without looking anything up (\cls{missing\_call}), claiming that a change was made without making
it (\cls{premature\_answer}), acting on a guessed value instead of asking for missing information
(\cls{guessed\_missing\_info}), and using \texttt{\$ref}s when program mode was not announced
(\cls{ref\_without\_capability}). Three concern effects and break the three rules of effect safety: performing the
destructive action at the point where the gold turn asks for confirmation (\cls{unconfirmed\_destructive}), repeating a
non-idempotent write (\cls{duplicate\_write}), and changing or deleting an entity by a guessed identifier without
reading it first (\cls{stale\_state\_write}). The last class, \cls{refusal}, applies a rule to public data instead of
world traces: at a public prefix whose gold action is a call, the bad action is a templated refusal such as ``I don't
have access to a tool that can do that'' (\S\ref{sec:refusal}). In total there are 19 classes. Eighteen are built on
EffectWorld traces, five by the optimizer and thirteen by rules; we call them \emph{world classes}. The nineteenth,
\cls{refusal}, is built on public data.

\subsection{Public data and training mixes}
\label{sec:data}

The SFT data mix combines gold EffectWorld trajectories with public tool-use trajectories, and it exists in two versions.
The \emph{raw mix} combines 5{,}976 gold world trajectories (2{,}988 training tasks, each in program and stepwise mode)
with 15{,}001 public trajectories sampled from the tool-calling and interactive subsets of Nemotron-Agentic-v1
\citep{nemotronagentic2025}, Hermes function calling \citep{hermesfc2024} and ToolACE \citep{liu2024toolace}, 20{,}977
examples in total. The \emph{clean mix} (\S\ref{sec:refusal}) drops 1{,}751 public trajectories that contain refusal
templates, caps first-turn no-call examples at 5\% of each public source and removes an arithmetic item from the world's
no-tool kind (\texttt{irrelevant} in Appendix~\ref{app:kinds}), leaving 20{,}975 examples. The pairs for the negative variants cap each of 17 world classes at 400
(6{,}580 pairs); the clean mix adds 1{,}000 \cls{refusal} pairs, for 7{,}580 (996 of the \cls{refusal} pairs survive
tokenisation).

\section{Experimental setup}
\label{sec:setup}

This section describes how the models are trained (\S\ref{sec:training}), the recipes that produce them
(\S\ref{sec:recipes}), the further backbones (\S\ref{sec:backbones}) and how they are evaluated (\S\ref{sec:eval}).

\subsection{Training objectives}
\label{sec:training}

Unless stated otherwise (\S\ref{sec:backbones}), all models share a frozen Qwen3-4B-Instruct-2507 backbone
\citep{yang2025qwen3} and train LoRA adapters
\citep{hu2022lora} with rank 16, $\alpha=16$ and dropout 0.05 on the attention projections and the up and down projections of the multilayer perceptron (MLP)
(26.0M trainable parameters), following LUNE's defaults: AdamW, 5\% warm-up and cosine decay, weight decay
0.01, gradient clipping at 1.0, bfloat16 precision, sequences up to 4{,}096 tokens.

Supervised fine-tuning is the first stage and the only one every model shares. The loss is the token-level negative
log-likelihood (NLL) of the assistant turns only. Qwen chat templates mark no generation spans,
so per-turn spans come from incremental rendering of the template. World trajectories appear in both program and
stepwise mode, so the same adapter serves both.

The second stage trains on pairs. A pair has a shared prefix $x$, a good action $y^+$ and a bad action $y^-$. The
\emph{bad span} $S^-$ is the part of $y^-$ to suppress: it starts at the first token where $y^-$ diverges from $y^+$ and
ends before the longest token suffix the two turns share. The closing tags and the end-of-turn token, which the good
turn needs as well, are therefore never suppressed. The one exception is a turn that stops before a required call, whose
premature end-of-turn token is its single target. With $p_\theta$ the adapted model and $\pi_{\mathrm{ref}}$ the SFT
adapter the variant starts from, held frozen as a second adapter, the UL and NPO losses are
\begin{align}
\mathcal{L}_{\mathrm{UL}} &= -\tfrac{1}{|S^-|}\sum_{t\in S^-}\log\bigl(1-p_\theta(y^-_t\mid x,y^-_{<t})\bigr),\\
\mathcal{L}_{\mathrm{NPO}} &= \tfrac{2}{\beta}\log\Bigl(1+\bigl(\pi_\theta(y^-\mid x)/\pi_{\mathrm{ref}}(y^-\mid x)\bigr)^{\beta}\Bigr),
\end{align}
where sequence probabilities are summed over the span as in the original formulation. Every pair step adds the NLL of
$y^+$ as a retain term, $\mathcal{L}=\mathcal{L}_{\mathrm{retain}}+\lambda\mathcal{L}_{\mathrm{neg}}$, and half of all
micro-batches are plain SFT batches. The pairwise variant uses DPO \citep{rafailov2023dpo} on the same pairs and the same
reference, with chosen and rejected log-probabilities over the whole good and bad turns (the shared prefix cancels in
the margin). A first round of variants was trained with three defects in these definitions; \S\ref{sec:variants} reports it as an ablation.

The comparison between objectives is controlled by construction. Every negative variant continues from the SFT adapter with a fresh optimizer for 300 steps at learning rate $5\times10^{-5}$,
$\lambda=0.3$ and $\beta=0.1$. The control continues SFT for the same 300 steps, so all variants see the same amount of
positive data and differ only in the negative term.

\subsection{Recipes}
\label{sec:recipes}

Every model in this paper is produced by one of the recipes in Table~\ref{tab:models}. Names with a subscript are SFT
recipes: they start from the base model and differ only in their training data, which the subscript describes. The
second-stage variants start from \Sclean{} and are named after their objective.

\begin{itemize}
  \item \textbf{Base} is Qwen3-4B-Instruct-2507 without an adapter.
  \item \textbf{\Sraw} is SFT on the raw mix for two epochs (1{,}269 steps of 32 sequences). It is the first model we
        trained and shows what the public data does on its own (\S\ref{sec:refusal}).
  \item \textbf{\Sclean} is the same recipe on the clean mix. It is the reference recipe: every other recipe below
        changes exactly one thing relative to it.
  \item \textbf{\Sdecl} replaces 10\% of the world training tasks (299 of 2{,}987) with tasks in which the tool the request
        needs has been removed from the tool list, so the only correct answer is a short decline that names the missing
        capability.
  \item \textbf{\Seff} keeps the clean mix but writes each tool's declared effect and idempotence into its description,
        in the style of the Model Context Protocol annotations, and renames the eighteen workspace tools in one of two
        synonym dialects for a third of the world trajectories (1{,}997 of 5{,}976), so that the confirmation habit
        cannot attach to a name.
  \item \textbf{\Snear} targets the cue \Sdecl{} missed (\S\ref{sec:bfcl-analysis}): a request offered tools from its own topic
        that cannot serve it. Candidates come from public data only: a public request whose first turn is answered by a
        call loses every tool it came with and is offered one to three tools from other public trajectories that are close
        to it in topic (cosine similarity of term frequency--inverse document frequency vectors of tool names and
        descriptions) but are not near-duplicates of the tool that was called.
        The base model answers each candidate; we keep those it declines and use its own answer as the target, so a
        candidate whose offered tool might serve the request after all is dropped. 1{,}500 of these replace as many public
        trajectories of the clean mix.
  \item The \textbf{second-stage variants} continue \Sclean{} for 300 steps with a fresh optimizer. The \emph{Control}
        continues SFT only; \emph{Unlikelihood} (UL), \emph{NPO} and \emph{DPO} add the objectives of
        \S\ref{sec:training} on the clean mix's pairs, with half of their micro-batches still plain SFT, so they differ
        from the control only in the negative term. Two \emph{ablations} rerun Unlikelihood with the pairs restricted by
        class.
\end{itemize}

The three other backbones of \S\ref{sec:backbones} are trained with the \Sclean{} and \Sdecl{} recipes unchanged.

\begin{table}[!ht]
\centering
\small
\caption{Recipes. Every adapter is LoRA rank 16 on the same frozen backbone. The second-stage variants start from
\Sclean{}'s adapter and share one schedule (300 steps, learning rate $5\times10^{-5}$, 32 sequences per step, half of
the micro-batches SFT, half pairs; $\lambda=0.3$, $\beta=0.1$).}
\label{tab:models}
\resizebox{\linewidth}{!}{%
\begin{tabular}{llll}
\toprule
Recipe & Starts from & Training data & Objective \\
\midrule
Base & -- & -- & -- \\
\midrule
\Sraw & Base & raw mix (20{,}977 examples) & NLL, 2 epochs \\
\Sclean & Base & clean mix (20{,}975 examples) & NLL, 2 epochs \\
\Sdecl & Base & clean mix, 10\% of world tasks are decline tasks (20{,}973) & NLL, 2 epochs \\
\Seff & Base & clean mix, effects in tool descriptions, a third of world tools renamed & NLL, 2 epochs \\
\Snear & Base & clean mix, 1{,}500 public examples replaced by near-miss declines & NLL, 2 epochs \\
\midrule
Control & \Sclean & clean mix & NLL, 300 steps \\
Unlikelihood (UL) & \Sclean & clean mix + 7{,}580 pairs (18 classes) & NLL + $\lambda\,\mathcal{L}_{\mathrm{UL}}$ \\
NPO & \Sclean & same & NLL + $\lambda\,\mathcal{L}_{\mathrm{NPO}}$, reference \Sclean \\
DPO & \Sclean & same & $\lambda$\,NLL + DPO on whole turns, reference \Sclean \\
UL without \cls{refusal} & \Sclean & clean mix + 6{,}580 pairs (17 world classes) & as Unlikelihood \\
UL, \cls{refusal} only & \Sclean & clean mix + 996 \cls{refusal} pairs & as Unlikelihood \\
\bottomrule
\end{tabular}}
\end{table}

\subsection{Backbones}
\label{sec:backbones}

To test whether the findings hold beyond one backbone, we also train three further small instruction-tuned models
from different lineages: Qwen3-1.7B, the same family at less than half the size (a hybrid reasoning model, evaluated with
thinking off), Phi-4-mini-instruct with 3.8B parameters, and Falcon3-3B-Instruct. Each is trained twice, with the \Sclean{}
recipe and with the \Sdecl{} recipe, using the hyperparameters of \S\ref{sec:training} and LoRA on the same
attention and MLP projections (Phi-4-mini fuses the query, key and value projections and the gate and up projections, so
its adapter also covers the gate). Every backbone is trained and evaluated in its own tool-call format, the one its BFCL
handler uses: Phi-4-mini receives the tool list in its system message and calls inside \texttt{<|tool\_call|>} blocks,
Falcon3 writes a JSON list of calls inside \texttt{<tool\_call>} tags. Falcon3's template inlines the tool list as indented
JSON, so about 15\% of the training examples exceed 4{,}096 tokens and are dropped (17{,}866 of 20{,}975 remain).

\subsection{Evaluation and compute}
\label{sec:eval}

All evaluations share one serving stack. World evaluations serve merged weights with vLLM \citep{kwon2023vllm}, render
prompts with the training template, decode greedily, allow 12 model turns, and run every task in program and stepwise
mode. NESTful (1{,}861 sequences) is scored structurally in program mode after mapping its variable references onto
\texttt{\$ref}s.

BFCL v4 is run on its three offline groups, 4{,}441 prompts in all, with the official handler for each model.
\emph{Non-live} prompts are written by the benchmark's authors and scored by comparing the abstract syntax tree of each
call with a reference. They come in four categories: \emph{simple} (one function; in Python, Java or JavaScript),
\emph{multiple} (one of several functions must be chosen), \emph{parallel} (several calls of one function) and
\emph{parallel-multiple}; \emph{irrelevance} prompts offer only functions that do not fit, and the correct answer is to
call none. \emph{Live} prompts were contributed by users and cover the same categories, plus \emph{relevance} prompts in
which a function does fit. \emph{Multi-turn} conversations run against stateful simulated back-ends, in a base setting
and with a missing function, missing parameters or a long context. We report BFCL's own accuracies per group.

Differences in success rates and accuracies are given in percentage points. For every difference between two models
quoted in the text we computed a 95\% paired bootstrap confidence interval (2{,}000 resamples), pairing the two models'
runs on the same task and mode or on the same BFCL prompt. To keep the text readable, these intervals are collected in
Tables~\ref{tab:diffs} and~\ref{tab:diffs2} (Appendix~\ref{app:diffs}); every difference listed there is significant, meaning its interval
excludes zero, unless the text marks it as not significant. Differences quoted without an interval, such as stepwise
gaps or per-kind changes, are descriptive. The intervals cover task sampling, not seed variation; \S\ref{sec:limitations}
reports a second training seed.

Every job runs on a single A100 40\,GB. An SFT recipe takes about 10.2 GPU-hours in three time-limited, self-resuming
slices (about 30\,s per step, 27.5\,GB peak). Each second-stage variant takes 2.5 to 4 GPU-hours, a world evaluation
about 10 minutes, and BFCL about 45 minutes.

\section{Results}
\label{sec:results}

Table~\ref{tab:overview} summarises every model on every benchmark. The subsections below first take the evaluation
sets one at a time, from the in-distribution test split to BFCL and NESTful, and then compare the models trained with
negatives (\S\ref{sec:variants}), the recipes that change the data (\S\ref{sec:v3}) and the other backbones
(\S\ref{sec:families}).

\begin{table}[t]
\centering
\small
\caption{Overview: every model on every benchmark. $\uparrow$: higher is better; $\downarrow$: lower is better.
EffectWorld columns give success in program / stepwise mode (\%); Avg is the mean over the three splits. Unconfirmed
destructive calls: calls that pay, cancel or refund without a confirmed proposal on the commerce split (547 tasks,
program mode). BFCL columns are accuracies (\%), Irrel.\ is irrelevance detection, and Avg is
the mean of the four BFCL columns. NESTful F1 is tool-name F1. Rows are grouped into the base model, the SFT recipes
(\S\ref{sec:recipes}), the second-stage variants continued from \Sclean{}, and two ablations of the unlikelihood
variant. Best value per column in bold. $^{*}$\Sraw{} declines irrelevance prompts because it refuses nearly every
prompt (\S\ref{sec:refusal}); this value is excluded when marking the best.}
\label{tab:overview}
\resizebox{\linewidth}{!}{
\begin{tabular}{lrrrrrrrrrrr}
\toprule
 & \multicolumn{4}{c}{EffectWorld success (prog / step) $\uparrow$} & \multirow{2}{*}{\shortstack{Unconfirmed destructive\\calls (commerce) $\downarrow$}} & \multicolumn{5}{c}{BFCL accuracy $\uparrow$} & NESTful \\
\cmidrule(lr){2-5}\cmidrule(lr){7-11}
Model & Test & Hard & Commerce & \textit{Avg} & & Non-live & Irrel. & Live & Multi-turn & \textit{Avg} & F1 $\uparrow$ \\
\midrule
Base & 52.6 / 50.6 & 51.7 / 53.2 & 74.4 / 74.8 & 59.6 / 59.5 & 7 & \textbf{88.0} & 88.8 & 76.2 & 21.9 & 68.7 & 0.39 \\
\midrule
\Sraw & 99.4 / \textbf{100.0} & 81.2 / 80.4 & 79.0 / 77.9 & 86.5 / 86.1 & 37 & 31.4 & 95.8$^{*}$ & 58.2 & 26.0 & 52.9 & 0.41 \\
\Sclean & 99.2 / \textbf{100.0} & 81.5 / 79.7 & 82.6 / 85.2 & 87.8 / \textbf{88.3} & 24 & 84.8 & 75.0 & 76.5 & 26.6 & 65.7 & 0.52 \\
\Sdecl & \textbf{99.6} / \textbf{100.0} & \textbf{83.7} / \textbf{83.9} & 77.9 / 79.5 & 87.1 / 87.8 & 42 & 87.0 & 75.0 & 77.7 & 26.4 & 66.5 & 0.53 \\
\Seff & 99.2 / \textbf{100.0} & 81.2 / 80.4 & 83.9 / 82.5 & 88.1 / 87.6 & 19 & 85.5 & 75.0 & 76.4 & 25.4 & 65.6 & \textbf{0.55} \\
\Snear & 99.0 / \textbf{100.0} & 83.0 / 81.5 & 73.9 / 75.3 & 85.3 / 85.6 & 64 & 86.7 & \textbf{89.6} & 77.9 & 25.6 & \textbf{70.0} & 0.51 \\
\midrule
Control & 99.4 / \textbf{100.0} & 81.3 / 80.7 & 83.7 / 84.1 & 88.1 / 88.2 & 24 & 85.5 & 77.5 & 76.3 & 24.0 & 65.8 & 0.50 \\
Unlikelihood & 99.0 / 95.6 & 77.8 / 71.4 & 62.9 / 66.5 & 79.9 / 77.8 & 16 & 85.5 & 66.2 & 77.9 & \textbf{27.8} & 64.4 & 0.43 \\
NPO & 99.0 / 99.2 & 77.4 / 75.3 & 71.3 / 76.0 & 82.6 / 83.5 & 16 & 84.6 & 58.3 & 75.2 & 24.8 & 60.7 & 0.47 \\
DPO & 99.4 / \textbf{100.0} & 78.7 / 78.1 & 67.1 / 70.4 & 81.7 / 82.8 & \textbf{5} & 84.5 & 69.2 & 76.3 & 24.2 & 63.6 & 0.52 \\
\midrule
UL without \cls{refusal} & 99.2 / 95.6 & 78.7 / 77.5 & 66.0 / 66.7 & 81.3 / 80.0 & 23 & 83.8 & 74.2 & 76.5 & 25.9 & 65.1 & 0.45 \\
UL, \cls{refusal} only & 99.2 / 98.0 & 78.7 / 77.5 & \textbf{87.8} / \textbf{86.5} & \textbf{88.6} / 87.3 & 16 & 86.4 & 36.2 & \textbf{79.4} & \textbf{27.8} & 57.5 & 0.49 \\
\bottomrule
\end{tabular}
}
\end{table}

\subsection{In-distribution test split}

On the in-distribution test split (Table~\ref{tab:overview}; full metrics in Table~\ref{tab:world}, Appendix~\ref{app:detailed}),
we compare the base model and \Sraw{}. The base model never uses
references in program mode, makes 94 unconfirmed destructive calls in 498 tasks (it asks before two thirds of the
deletions but rarely before sending or moving), and changes unrelated entities in 10\% of tasks. \Sraw{} solves 99.4\% of tasks in program mode and all of them stepwise, with no excess calls, no violations
and 60\% fewer output tokens. Program mode reaches the same success as stepwise with 41\% fewer action turns, 25\% fewer
model calls and 23\% fewer prompt tokens. \Sclean{}, trained on the clean mix, is
indistinguishable here (99.2\% and 100\%). Because both saturate this split, it cannot rank the negative variants.

\subsection{Hard split}

On the hard split (Table~\ref{tab:hard}), \Sraw{} raises success from 52\% to 80--81\%, a gain of 29.4 points in
program mode. It removes almost all effect-safety violations (0.15 to 0.003 per task) and, when the user says there is
no need to ask before cancelling, it now proceeds (32\% to 96\%). It fails in three places that its training data never
rewarded. It never asks which of two same-named contacts is meant: success on both ambiguous-contact kinds stays at 0,
and when scheduling a meeting it invites the wrong person in 22 of 45 tasks. It acts on entities that do not exist,
replying to a sender who never wrote in 29 of 47 missing-sender tasks, where the more hesitant base model did better.
And it deletes the wrong one of two same-titled events in 12 of 29 cases. These failures belong to the guessed-information,
premature-answer and wrong-argument-value classes that the negative variants target. \Sclean{} behaves the same overall
(81.6\% and 79.7\%, 29.8 points above the base in program mode). It no longer hesitates on missing events (72\% to 100\%)
but acts even more readily on missing senders (38\% to 19\% success).
The variant columns are discussed in \S\ref{sec:variants}.

\begin{table}[t]
\centering
\small
\caption{Held-out hard split (775 tasks, 24 tools including six distractors). Success per kind in program mode; ``new''
marks kinds absent from training. Control, UL, NPO and DPO are the variants continued from \Sclean{} (\S\ref{sec:variants}). Best value per row in bold (ties included, none when all are equal); italic rows are means over the kinds above them.}
\label{tab:hard}
\resizebox{\linewidth}{!}{
\begin{tabular}{lrrrrrrrrrr}
\toprule
 & Base & \Sraw & \Sclean & \Sdecl & \Seff & \Snear & Control & UL & NPO & DPO \\
\midrule
Overall, program mode & 51.7 & 81.2 & 81.5 & \textbf{83.7} & 81.2 & 83.0 & 81.3 & 77.8 & 77.4 & 78.7 \\
Overall, stepwise & 53.2 & 80.4 & 79.7 & \textbf{83.9} & 80.4 & 81.5 & 80.7 & 71.4 & 75.3 & 78.1 \\
Effect-safety violations per task (prog) & 0.15 & 0.003 & 0.004 & \textbf{0.000} & 0.008 & 0.004 & 0.004 & 0.043 & 0.001 & \textbf{0.000} \\
Output tokens per task (prog) & 275 & 109 & 115 & \textbf{107} & 110 & 110 & 118 & 134 & 116 & 115 \\
\midrule
Ambiguous contact, meeting (new) & 0.0 & 0.0 & 0.0 & 0.0 & 0.0 & 0.0 & 0.0 & 0.0 & 0.0 & 0.0 \\
Ambiguous contact, unread mail (new) & 0.0 & 0.0 & 0.0 & 0.0 & 0.0 & 0.0 & 0.0 & 0.0 & 0.0 & 0.0 \\
Missing event (new) & \textbf{100.0} & 72.3 & \textbf{100.0} & 97.9 & 93.6 & \textbf{100.0} & \textbf{100.0} & 59.6 & 74.5 & 85.1 \\
Missing sender (new) & \textbf{74.5} & 38.3 & 19.1 & 57.5 & 25.5 & 51.1 & 17.0 & 2.1 & 6.4 & 25.5 \\
No open tasks (new) & \textbf{100.0} & \textbf{100.0} & \textbf{100.0} & \textbf{100.0} & \textbf{100.0} & \textbf{100.0} & \textbf{100.0} & \textbf{100.0} & 97.9 & 74.5 \\
Same-title events, dated (new) & 51.7 & 58.6 & 58.6 & 55.2 & \textbf{62.1} & 51.7 & \textbf{62.1} & 48.3 & 55.2 & 51.7 \\
Cancel, ``no need to ask'' (new) & 31.9 & 95.7 & \textbf{100.0} & \textbf{100.0} & \textbf{100.0} & 97.9 & \textbf{100.0} & \textbf{100.0} & 97.9 & 97.9 \\
Near-domain, no tool needed (new) & \textbf{100.0} & \textbf{100.0} & \textbf{100.0} & \textbf{100.0} & 89.4 & \textbf{100.0} & 95.7 & \textbf{100.0} & 97.9 & \textbf{100.0} \\
\textit{Average, eight new kinds} & 57.3 & 58.1 & 59.7 & \textbf{63.8} & 58.8 & 62.6 & 59.4 & 51.2 & 53.7 & 54.3 \\
Nine training kinds, range & 8.5--100.0 & 100.0 & 97.9--100.0 & 97.9--100.0 & 97.9--100.0 & 95.7--100.0 & 97.9--100.0 & 95.7--100.0 & 83.0--100.0 & 95.7--100.0 \\
\textit{Average, nine training kinds} & 46.3 & \textbf{100.0} & 99.3 & 99.5 & 99.5 & 99.1 & 99.3 & 99.5 & 96.9 & 98.6 \\
\bottomrule
\end{tabular}
}
\end{table}

\subsection{Commerce split}

The commerce split (547 tasks, 17 tools never seen in training, \S\ref{sec:splits}; Table~\ref{tab:overview}, with full metrics in Table~\ref{tab:commerce}) tests whether
the effect discipline transfers to tools the model has never seen. \Sraw{} and \Sclean{} raise success from 74\% to 79\% and 83\% (85\% stepwise; for \Sclean{} a gain of 8.2 points over the base) and cut excess calls from 0.86 to 0.46 and 0.63 per task and output tokens from
336 to 108 and 121, without a single training example of these tools. Effect safety, however, transfers only
partly. The base model asks before paying, cancelling or refunding on these tools (7 unconfirmed destructive calls in
547 tasks), the adapters make 24--37, and two kinds end up worse than the base. On buying an item (75\%
to 21--39\% success), every adapter pays without asking on 13--19 of 28 tasks, and when the user pushes back on the wording of
the answer it pays again (Figure~\ref{fig:trace-models}). Changing a delivery address needs the order to be read first
(88\% to 38--45\%): the adapters skip the read.

\subsection{BFCL}

Table~\ref{tab:bfcl} reports BFCL v4. \Sraw{} loses 57 points of non-live accuracy and 18 points of live accuracy,
while gaining 7 points of non-live irrelevance detection and 9.5 points on multi-turn base; \S\ref{sec:refusal}
explains this pattern. \Sclean{}, trained on the clean mix, recovers to within 3.3 points of the base on non-live
accuracy. On simple Python calls it is level with the base (95.3 against 95.5, not significant); the remaining gap lies in
simple Java and JavaScript calls and in parallel-multiple prompts, where it is 6.0 points lower. It matches the base on
live accuracy (76.5) and leads on multi-turn (26.6 against 21.9, and 9.5 points higher on multi-turn base). The price is
irrelevance detection, which falls from 88.8 to 75.0 on non-live prompts and from 81.1 to 74.4 on live ones. The negative variants move the same dial further
(\S\ref{sec:variants}); \S\ref{sec:bfcl-analysis} explains where fine-tuning gains on BFCL and where it loses.

\begin{table}[t]
\centering
\small
\caption{BFCL v4 offline groups (accuracy, \%). Overall scores across all v4 groups are not reported because we do not
run the web-search and memory groups. Simple: BFCL's mean over Python, Java and JavaScript simple prompts. Irrel.\ and
rel.: irrelevance and relevance detection. Best value per column in bold; Avg is the mean of the nine columns (the BFCL avg of Tables~\ref{tab:overview}
and~\ref{tab:families} averages four).}
\label{tab:bfcl}
\resizebox{\linewidth}{!}{
\begin{tabular}{lrrrrrrrrrr}
\toprule
Model & Non-live & Simple & Multiple & Parallel & Irrel. & Live & Live irrel. & Live rel. & Multi-turn & \textit{Avg} \\
\midrule
Base & \textbf{88.0} & \textbf{75.2} & 94.5 & \textbf{92.5} & 88.8 & 76.2 & 81.1 & \textbf{87.5} & 21.9 & 78.4 \\
\Sraw & 31.4 & 19.6 & 7.0 & 44.0 & \textbf{95.8} & 58.2 & \textbf{87.6} & 62.5 & 26.0 & 48.0 \\
\Sclean & 84.8 & 69.1 & 95.5 & 90.5 & 75.0 & 76.5 & 74.4 & \textbf{87.5} & 26.6 & 75.5 \\
\Sdecl & 87.0 & 74.7 & 95.5 & \textbf{92.5} & 75.0 & 77.7 & 66.4 & \textbf{87.5} & 26.4 & 75.9 \\
\Seff & 85.5 & 72.3 & 95.0 & 90.5 & 75.0 & 76.4 & 71.7 & 75.0 & 25.4 & 74.1 \\
\Snear & 86.7 & \textbf{75.2} & 95.0 & 92.0 & 89.6 & \textbf{77.9} & 77.6 & \textbf{87.5} & 25.6 & \textbf{78.6} \\
Control & 85.5 & 68.8 & \textbf{96.5} & \textbf{92.5} & 77.5 & 76.3 & 73.0 & 81.2 & 24.0 & 75.0 \\
Unlikelihood & 85.5 & 72.9 & 95.5 & 91.0 & 66.2 & \textbf{77.9} & 62.0 & \textbf{87.5} & \textbf{27.8} & 74.0 \\
NPO & 84.6 & 74.2 & 95.0 & 89.0 & 58.3 & 75.2 & 62.1 & \textbf{87.5} & 24.8 & 72.3 \\
DPO & 84.5 & 71.9 & 95.5 & 88.5 & 69.2 & 76.3 & 70.7 & \textbf{87.5} & 24.2 & 74.3 \\
\bottomrule
\end{tabular}
}
\end{table}

\subsection{NESTful}

Base and \Sraw{} are near the floor on NESTful: exact sequence match 0.0\% and 0.1\%, tool-name F1 0.39 and 0.41,
per-call argument match 15.3\% and 14.6\%. \Sclean{} lifts name F1 to 0.52 and argument match to 17.3\% at 55 output
tokens per sequence (63\% fewer than the base); the variants trained with negatives (\S\ref{sec:variants}) add nothing to this (F1 0.43--0.50). All models rarely chain
results through \texttt{\$ref}s here, because the world only teaches references to identifiers and addresses, not to
computed values.

\subsection{Variants trained with negatives}
\label{sec:variants}

Every variant continues from \Sclean{} for 300 steps with the protocol of \S\ref{sec:training}; the control continues
SFT only, so the variants differ from it only in the negative term. Table~\ref{tab:overview} summarises the comparison,
and Tables~\ref{tab:hard}, \ref{tab:bfcl} and \ref{tab:commerce} give the detail. Differences below are against the
control unless stated otherwise.

In distribution, the objectives do no harm. Every variant stays at or above 99\% on the test split, and the validation
NLL of the positives stays within $1.4\times$ the control's.

On the hard split they do not fix the failures they target. No variant asks which of two same-named contacts is meant
(0\% on both ambiguous-contact kinds). Both suppression variants act on non-existent senders more often than the control
(2--6\% success against 17\%), and the kind with two same-titled events does not move. Overall hard-split success falls
by 3.5 points under unlikelihood and by 3.9 under NPO in program mode, and by 9.3 and 5.3 points stepwise.

On the unseen domain suppression costs 20.8 points under unlikelihood and 12.4 under NPO. The losses concentrate where a suppressed class has a close look-alike. The \cls{unbatched}
class penalises adding items one at a time, and the kind that adds several items to the cart falls from 96\% to 30\%
under unlikelihood. Reporting that a requested product does not exist falls from 72\% to 28\% and 46\%.

What the negatives do change is how readily the model calls. Unlikelihood gains 1.6 points of BFCL live accuracy and
3.8 points of multi-turn accuracy, most clearly on multi-turn base (5.0 points) and live multiple (2.2 points). Its
irrelevance detection falls from 77.5 to 66.3, and NPO's to 58.3. Of the 240
irrelevance prompts, the control declines 186, unlikelihood 159 and NPO 140 (Table~\ref{tab:nocall}). The \cls{refusal}
class, which contrasts a call with a templated refusal, is the largest class (13\% of all pairs), and its direction is
the one the model follows. The control's own 300 SFT steps change nothing measurable relative to \Sclean{}: 0.3 points lower on
the hard split, 0.4 higher on commerce and 4.0 lower on BFCL multi-turn base, none of them significant.

DPO, the pairwise objective, is the most careful variant but not the most successful. It makes no effect-safety
violation on the hard split, 5 unconfirmed destructive calls on commerce against the control's 24, and fewer excess
calls everywhere. Its success is nonetheless lower, by 2.6 points on the hard split and 16.6 on commerce. On BFCL
it gains nothing on live or multi-turn prompts and loses 8.3 points of irrelevance detection.

Two ablations of the unlikelihood variant locate the cause (last rows of Table~\ref{tab:overview}). Without the
\cls{refusal} class, irrelevance detection recovers most of the way, from 66.2 under unlikelihood with all classes to
74.2, which is 3.3 points below the control. The live and multi-turn gains disappear: multi-turn base accuracy is 7.0
points lower than with all classes. The losses against the control remain, at 2.6 points on the hard split and 17.7 on
commerce. With \emph{only} the \cls{refusal} class, the model calls on almost everything: irrelevance detection falls to
36.2, 41.2 points below the control, while live-multiple and multi-turn accuracy rise by 3.2 and 3.8 points. On the
commerce split, where every kind but one needs calls, this is the only variant above its control, by 4.0 points. The public-prefix refusal pairs are
therefore a knob for how readily the model calls, and the seventeen optimizer- and rule-derived classes, the ones this
paper motivates, account for the losses and for no gain.

One earlier result belongs here because it explains how the objectives above were calibrated. The same variants were
first trained from \Sraw{} with three defects in the objectives of \S\ref{sec:training}:
\begin{enumerate}
  \item The suppression span ran to the end of the turn, so the closing tag and the end-of-turn token, which the good
        turn needs as well, were suppressed too.
  \item $\beta=0.1$ was applied to per-token mean log-probabilities instead of sums, so the loss kept pushing the span
        down until it was about 30 nats per token below the reference.
  \item The reference was the backbone. The SFT adapter already gives the bad spans about 7 nats per token less
        log-probability than the backbone does, so a correctly scaled loss would have been saturated at step 0.
\end{enumerate}
Every negative variant lost to its control (hard split 36.6--78.6\% against 82.6\%; commerce 15.9--65.1\% against
80.3\%). NPO degenerated into repeated tokens in 48\% of hard-split tasks, and NPO and DPO left 29--68\% of tool-call
blocks unclosed. We report this round because its failure, a model taught not to close its calls, is invisible in the
training curves, where the negative loss went to zero, and appears only in held-out rollouts.

\subsection{Recipes that change the data}
\label{sec:v3}

Two findings above point at data, not objectives: the adapters decline less than the base on BFCL's irrelevance prompts
(75.0 against 88.8) because the refusal filter of the clean mix removed the examples that taught declining, and they confirm less
reliably on unseen tools (24 against 7 unconfirmed destructive calls on commerce) because the effect declarations never
reach the model. The \Sdecl{} and \Seff{} recipes (\S\ref{sec:recipes}) each change
one thing in \Sclean{} to address one of them: \Sdecl{} adds tasks whose only correct answer is a decline because the
needed tool is missing, and \Seff{} shows the effect declarations in the tool descriptions while renaming the tools, so
that the confirmation habit cannot attach to a name.

Neither of these two levers reaches the base on its own target; a third, \Snear{}, motivated by the analysis of
\S\ref{sec:bfcl-analysis}, is discussed at the end of this subsection. \Sdecl{} leaves BFCL irrelevance detection exactly where \Sclean{} had it
and lowers it by 8.0 points on live prompts. A ``tool absent, so
decline'' positive does not transfer to BFCL's cue, where the functions are present and the request does not match them
(\S\ref{sec:bfcl-analysis}). It does generalise in a direction we had not aimed at. On the hard split it is the best
model so far, 2.2 points above \Sclean{} in program mode and 4.1 stepwise, with no effect-safety violation, chiefly
because it stops acting on non-existent senders (19\% to 58\% success). It is also the closest to the base on
non-live BFCL accuracy (87.0 against 88.0, and 1.2 points above \Sclean). On the commerce split, however, it loses 4.8
points. It makes 0.37 fewer calls per task but confirms less often (24 to 42
unconfirmed destructive calls), as if ``decline'' had generalised to ``do less''. \Seff{} moves confirmation transfer in the intended direction
but not far: 24 to 19 unconfirmed destructive calls (fewer violations per task, but not significantly)
and purchases made after a confirmation from 32\% to 50\%. Nothing else changes except a loss of 5.0 points on BFCL
multi-turn base and the best NESTful name F1 (0.55). Rendering the annotations helps only when the training data also shows destructive tools
under many names; a third of the trajectories in two dialects is evidently too little, and the base model's habit of
asking before any purchase remains the stronger prior. Hammer \citep{lin2025hammer} reports the same attachment of
function-calling behaviour to naming conventions and counters it by masking function names during training.

\Snear{} supplies the cue that \Sdecl{} missed: an offered tool from the request's own topic that cannot serve it. Of the
6{,}000 candidates, the base model declined 3{,}831, and 1{,}500 of these replaced public trajectories. It is the first
recipe to match the base model on BFCL irrelevance detection: 89.6 against 88.8 for the base, a difference that is not
significant, and 75.0 for \Sclean{}, a gain of 14.6 points. It declines 215 of the 240 irrelevance prompts, the base 213
(Table~\ref{tab:nocall}). Live irrelevance detection rises by 3.2 points but stays 3.5 points below the base. Non-live
and live accuracy rise as well, to 86.7 and 77.9, so its BFCL average of 70.0 is the only one above the base's 68.7
(Table~\ref{tab:overview}). On the hard split it gains 1.4 points and, like \Sdecl{}, stops acting on non-existent
senders (19\% to 51\% success). The price is again paid on the unseen domain. Commerce success falls by 8.8 points, back
to the base's level, because the model now refunds and buys without asking: unconfirmed destructive calls rise from 24
to 64, refunds of delivered orders fall from 86\% to 5\% success and purchases from 32\% to 7\%. Contrasting ``the
offered tool does not fit, so decline'' with ``it fits, so call'' appears to sharpen the decision to call at the expense
of the confirmation step that should come between the two.

\subsection{Other backbones}
\label{sec:families}

Table~\ref{tab:families} compares the base models and recipes of the three backbones of \S\ref{sec:backbones} with
Qwen3-4B. The base models leave far more room than Qwen3-4B does. The best of them solves about a
quarter of the in-distribution tasks. Qwen3-1.7B is close to its larger sibling on BFCL non-live and live accuracy but
solves 25\% of the world test, 20\% of the hard split and 13\% of the commerce split, with 0.44 tool errors per task.
Phi-4-mini rarely emits a tool call in its native format: in EffectWorld it makes 0.15 to 0.34 calls per task. On BFCL
the official decoder finds a call in none of its answers. 86\% of them are prose, typically an explanation with the call
written as a Python code block, and the rest are untagged JSON or calls opened with a doubled \texttt{<|tool\_call|>}
tag that the decoder cannot parse. It therefore scores 0 on every category that needs a call and 100 on irrelevance, because BFCL counts any
answer without a parseable call as a correct decline. To
separate format from ability we re-scored its answers with BFCL's own checker after a lenient decoding step that also
reads calls written as code, untagged JSON and doubled tags. The base model then expresses a call in 60 to 68\% of the
prompts that need one and reaches 24.7 non-live and 39.2 live accuracy. All reported numbers use the official decoder,
so that the fine-tuned model is judged in the same way, and our own parser in EffectWorld accepts all three forms. Falcon3 calls readily but wastefully, with 1.75
excess calls per task on the test split, and declines only 35\% of BFCL's irrelevance prompts.
SFT carries over to every family in distribution and on the hard split, less so to the unseen domain. \Sclean{}
lifts in-distribution success to 99\% for Qwen3-1.7B and Falcon3-3B and to 69\% for Phi-4-mini, whose remaining
failures are wrong plans rather than format errors, such as an email sent without a recipient. On the hard split it adds
54.7, 32.0 and 52.1 points for Qwen3-1.7B, Phi-4-mini and Falcon3-3B, and on commerce 41.0, 25.8 and 27.2 points. Commerce
success nonetheless stays between 35\% and 54\%, far below Qwen3-4B's 83\%: the smaller the model, the less of the
training domain carries over to unseen tools. The confirmation habit fails to transfer in every family. On commerce the
three base models make 1, 2 and 2 unconfirmed destructive calls and their \Sclean{} adapters 45, 6 and 39, the same
pattern as Qwen3-4B (7 against 24).

On BFCL the families diverge. Phi-4-mini gains most, from 0 to 74.5 non-live and 63.8 live accuracy, and the lenient re-scoring shows
where the gain comes from. About a third of the non-live gain is format alone, since the base model reaches 24.7 once its
answers are decoded leniently. The rest comes from two behaviours that fine-tuning teaches. The model now answers 86 to
97\% of call prompts with a call, against 60 to 68\% even under lenient decoding, so it stops answering in prose. And it
gets the structure of the calls right: parallel and parallel-multiple accuracy rise from 12.0 and 6.5 under lenient
decoding to 78.0 and 74.5, because the base model writes one call or a loop where several calls are needed. Its
irrelevance detection falls from 87.5 under lenient decoding to 73.8, the same shift towards calling seen in the other
families. Falcon3-3B
gains 26.5 points of non-live and 12.7 of live accuracy and keeps its low irrelevance detection (34.2 against 34.6, not
significant). Qwen3-1.7B loses: its irrelevance detection collapses from 81.2 to 35.4 and its live irrelevance detection
from 73.4 to 23.8, the effect of the refusal filter (\S\ref{sec:refusal}) at a much larger scale, and its live accuracy
falls from 74.5 to 67.1.

\Sdecl{} repeats on every family what it showed on Qwen3-4B, more strongly the smaller the model: declining generalises
to doing less. It cuts effect-safety violations on commerce to 0.026, 0.009 and 0.000 per task and raises irrelevance
detection by 8.3 and 5.0 points on Qwen3-1.7B and Phi-4-mini, but it does so by calling less. On commerce Phi-4-mini's
calls per task fall from 1.96 to 1.01 and Falcon3-3B's from 6.20 to 0.01, so Falcon3-3B with this recipe almost never
calls an unseen tool and its commerce success falls from 35.4\% to 17.4\%. Phi-4-mini also loses 29.3 points in
distribution, and Falcon3-3B 21.5 points on the hard split.

\begin{table}[t]
\centering
\small
\caption{Other backbones: base models and the two SFT recipes. World columns: success in program mode (\%); Viol./task:
effect-safety violations per task, mean over the three splits; BFCL irrel.: irrelevance detection; BFCL avg: mean of the four BFCL columns. The Qwen3-4B rows
repeat \Sclean{} and \Sdecl{} from the earlier tables. Best value per backbone in bold. $^\dagger$BFCL's decoder finds no call
in any of the Phi-4-mini base model's answers, which are prose, code blocks, untagged JSON or doubled tags; with lenient
decoding it reaches 24.7 non-live and 39.2 live accuracy (\S\ref{sec:families}). $^{*}$BFCL counts any answer without a
parseable call as a correct decline, so this 100 does not mean the model declines; with lenient decoding it is 87.5. Marked cells are excluded when
marking the best value.}
\label{tab:families}
\resizebox{\linewidth}{!}{
\begin{tabular}{llrrrrrrrrrrr}
\toprule
Backbone & Model & Test & Hard & Commerce & \textit{World avg} & Viol./task & BFCL non-live & BFCL irrel. & BFCL live & BFCL multi-turn & \textit{BFCL avg} & NESTful F1 \\
\midrule
\multirow{3}{*}{Qwen3-4B-2507} & Base & 52.6 & 51.7 & 74.4 & 59.6 & 0.12 & \textbf{88.0} & \textbf{88.8} & 76.2 & 21.9 & \textbf{68.7} & 0.39 \\
 & \Sclean & 99.2 & 81.5 & \textbf{82.6} & \textbf{87.8} & \textbf{0.032} & 84.8 & 75.0 & 76.5 & \textbf{26.6} & 65.7 & 0.52 \\
 & \Sdecl & \textbf{99.6} & \textbf{83.7} & 77.9 & 87.1 & 0.034 & 87.0 & 75.0 & \textbf{77.7} & 26.4 & 66.5 & \textbf{0.53} \\
\midrule
\multirow{3}{*}{Qwen3-1.7B} & Base & 25.3 & 20.4 & 12.6 & 19.4 & 0.039 & \textbf{84.2} & \textbf{81.2} & \textbf{74.5} & \textbf{13.5} & \textbf{63.4} & 0.43 \\
 & \Sclean & \textbf{99.0} & 75.1 & \textbf{53.6} & \textbf{75.9} & 0.046 & 80.4 & 35.4 & 67.1 & 11.5 & 48.6 & \textbf{0.53} \\
 & \Sdecl & 98.8 & \textbf{75.9} & 45.9 & 73.5 & \textbf{0.018} & 80.6 & 43.8 & 67.1 & 10.8 & 50.5 & 0.50 \\
\midrule
\multirow{3}{*}{Phi-4-mini (3.8B)} & Base & 11.2 & 20.0 & 13.9 & 15.0 & 0.007 & 0.0$^\dagger$ & 100.0$^{*}$ & 0.0$^\dagger$ & 0.0$^\dagger$ & 25.0$^\dagger$ & 0.14 \\
 & \Sclean & \textbf{68.9} & \textbf{52.0} & \textbf{39.7} & \textbf{53.5} & 0.031 & \textbf{74.5} & 73.8 & \textbf{63.8} & \textbf{7.4} & 54.9 & \textbf{0.45} \\
 & \Sdecl & 39.6 & 40.4 & 34.5 & 38.2 & \textbf{0.003} & 73.8 & \textbf{78.8} & 62.0 & 7.2 & \textbf{55.4} & 0.40 \\
\midrule
\multirow{3}{*}{Falcon3-3B} & Base & 3.8 & 12.7 & 8.2 & 8.2 & 0.074 & 53.2 & 34.6 & 54.7 & 1.0 & 35.9 & 0.46 \\
 & \Sclean & \textbf{99.0} & \textbf{64.8} & \textbf{35.4} & \textbf{66.4} & 0.043 & \textbf{79.7} & 34.2 & \textbf{67.4} & \textbf{5.9} & 46.8 & \textbf{0.49} \\
 & \Sdecl & 98.4 & 43.2 & 17.4 & 53.0 & \textbf{0.000} & 79.1 & \textbf{37.9} & 67.3 & 5.4 & \textbf{47.4} & 0.46 \\
\bottomrule
\end{tabular}
}
\end{table}

\section{Analysis}
\label{sec:analysis}

\subsection{Public data teaches refusal}
\label{sec:refusal}

\Sraw{} answers 374 of 400 BFCL simple-Python prompts without a call; the base model calls on all 400. The pattern holds
across categories (Table~\ref{tab:nocall}). In 255 of the 374 cases the answer says the model does not have access to the
tool it was just given; most of the rest compute the answer in prose. Gains in irrelevance detection and multi-turn base
are the same shift seen from the other side: \Sraw{} has moved its prior towards not calling.

\begin{table}[t]
\centering
\small
\caption{Prompts answered without any tool call in BFCL v4. The upper rows need a call, so fewer is better; the lower
two should be declined, so more is better. Best value per row in bold.}
\label{tab:nocall}
\resizebox{\linewidth}{!}{
\begin{tabular}{lrrrrrrrrrrr}
\toprule
Category & Prompts & Base & \Sraw & \Sclean & \Sdecl & \Seff & \Snear & Control & UL & NPO & DPO \\
\midrule
Simple (Python) & 400 & \textbf{0} & 374 & \textbf{0} & \textbf{0} & \textbf{0} & \textbf{0} & \textbf{0} & \textbf{0} & \textbf{0} & \textbf{0} \\
Multiple & 200 & 2 & 186 & 2 & \textbf{0} & 1 & 1 & 1 & \textbf{0} & \textbf{0} & 1 \\
Parallel & 200 & \textbf{0} & 103 & \textbf{0} & \textbf{0} & \textbf{0} & \textbf{0} & \textbf{0} & \textbf{0} & \textbf{0} & \textbf{0} \\
Live simple & 258 & 10 & 89 & 4 & \textbf{2} & \textbf{2} & 7 & 4 & \textbf{2} & \textbf{2} & \textbf{2} \\
Live multiple & 1{,}053 & 21 & 201 & 25 & 8 & 25 & 12 & 20 & \textbf{3} & 6 & 16 \\
\midrule
Irrelevance (should decline) & 240 & 213 & \textbf{230} & 180 & 180 & 180 & 215 & 186 & 159 & 140 & 165 \\
Live irrelevance (should decline) & 884 & 713 & \textbf{773} & 656 & 586 & 632 & 685 & 644 & 544 & 544 & 621 \\
\bottomrule
\end{tabular}
}
\end{table}

The cause is in the public part of the mix. About 780 sampled Nemotron trajectories contain refusal phrases such as ``I
don't have access to tools that\ldots'' or ``my current capabilities are limited to the tools listed'', some with the
refusal copied into the user turn. About 16\% of all examples in the raw mix open without a call, and the world's no-tool kind
included an arithmetic question answered in prose. A 4B model with a small adapter generalises these into a refusal
template for any computation-style request. The mechanism resembles the exaggerated safety of chat models, which refuse
safe prompts that share surface features with unsafe ones \citep{rottger2024xstest}; here the shared feature is the
shape of a request for a computation. The in-domain world metrics give no warning, because every world task either
needs a call or is plainly off-topic.

The clean mix removes this signal at the source and adds explicit \cls{refusal} negatives. It keeps 7.3\% first-turn no-call
examples, most of them world tasks that legitimately start by asking. \Sclean{} calls on every simple prompt again
(Table~\ref{tab:nocall}), but it also calls on 33 more of the 240 irrelevance prompts than the base: the filter removes
the examples that taught declining together with the ones that taught refusing. The \cls{refusal} negatives, which
contrast a call against a templated refusal on public prefixes, push further in the same direction (159, 140 and 165
declined prompts under unlikelihood, NPO and DPO), so they add no discrimination beyond the filter. Recovering the base's
irrelevance detection would need the opposite contrast as well, declining when appropriate against calling, which none
of our negative classes provides. The declining positives of \Sdecl{} are not that contrast either: they decline because a
tool is missing, and \Sdecl{} declines exactly as often as \Sclean{} on the irrelevance prompts (180 of 240) and less often on
the live ones (586 against 656 of 884).

\subsection{When fine-tuning beats the base model on BFCL, and when it does not}
\label{sec:bfcl-analysis}

Only \Snear{} beats the base model on the BFCL average, and it was designed after the analysis below. For the other
recipes the average hides two opposite movements. Counted per
prompt over the single-turn categories, \Sclean{} answers 157 prompts correctly that the base model gets wrong and 272
wrongly that the base model gets right. Half of the 272 (36 non-live and 98 live) are irrelevance prompts on which it
calls a function instead of declining. The remainder are spread thinly over the call categories.

The gains come where BFCL rewards something EffectWorld teaches. On multi-turn base \Sclean{} solves 35.5\% of
conversations against the base model's 26.0\%, on the turns that withhold a required parameter 22.5\% against 15.5\%,
and on live simple calls 82.6\% against 78.3\%. These are the behaviours of the world's tasks: carrying state across
turns, asking for information that is missing instead of inventing it, and calling directly when one function fits.

The irrelevance losses have one shape. In the failures we sampled, the model calls the function closest to the
request in topic and fills the arguments the request does not provide, the failure that \citet{wang2025learningtoask}
attribute to next-token training. Asked for an \emph{external} rate of return, it
calls a tool that computes the internal one; asked for compound interest, it calls a mortgage calculator; asked for a
lion's dominant genetic trait, it passes \texttt{"Lion"} as a variant identifier to a variant-frequency look-up. The
training data never shows this situation. Every one of the 14{,}999 public trajectories in the clean mix contains a call.
The world's no-call tasks are general requests (the no-tool training kind) or workspace-flavoured requests that ask for
no action (the near-domain kind of the hard split, which \Sclean{} declines without exception); in neither does an
offered tool look applicable without being so. A request that matches a tool's topic but not its function is therefore always
answered by a call in training, and the base model's calibration on such prompts, presumably learnt from data of this kind,
is overwritten. \Sdecl{} confirms that the cue matters: its declines are triggered by a tool that is absent, and on BFCL,
where a function with a plausible name is always present, it declines no more often on non-live irrelevance prompts and
less often on live ones (74.4 to 66.4). \Snear{} supplies the missing cue directly, and it works: irrelevance detection rises from 75.0 to 89.6, level with
the base model, although the same contrast costs confirmation on unseen tools (\S\ref{sec:v3}). The cue is the
limited-functionality setting of ToolBeHonest
\citep{zhang2024toolbehonest}, and When2Call \citep{ross2025when2call} reports that a preference objective over call,
ask and decline options teaches this decision better than SFT, a contrast with our negatives worth testing on this
cue.

The remaining gap on calls is small, and little of it is about effects. The base model already scores 95.5 on simple
Python calls and 94.5 on multiple-function prompts, so there is little room above it. The rest of the losses are on
parallel-multiple prompts (90.0 to 84.0, mostly a wrong number of calls), on Java and JavaScript prompts whose argument
types our JSON-only data never shows (64.0 to 60.0 and 66.0 to 52.0; \Sdecl{} recovers JavaScript), and on live prompts
whose expected value is a canonical form of what the user wrote (\texttt{Chicago, IL} where the user wrote
\texttt{Chi-town}). The effect-safety training is not the cause: in only 5 of the 122 lost prompts of the Python and live
call categories does the model answer with a question instead of a call. Finally, BFCL scores the form of a call, not
whether it should have been made with confirmation, whether it repeats an earlier call, or whether it writes over state
the model has not read; on those properties, which are what the effect declarations are for, the comparison is the one of
Tables~\ref{tab:overview} and~\ref{tab:hard}.

\section{Limitations}
\label{sec:limitations}

EffectWorld is small: nine task kinds in one training domain, one test-only domain, and template utterances
(paraphrases generated by a larger language model are not yet used). The scripted user never changes its mind and
confirms every proposal it recognises as one, including a proposal to act on the wrong entity; the verifier counts any
proposal the scripted user confirmed as confirmed. A model that asks, is told the entity does not exist, and then
proposes a substitute is therefore scored as if it had not asked (Figure~\ref{fig:trace-models}). Efficiency negatives are only as good as the optimizer's provenance heuristics; legality is checked by replay of the
recorded trace, not proved for all states. BFCL multi-turn and $\tau^2$-bench, which test policies with a model-based
user, are not yet evaluated end to end. The detailed analyses use one backbone, Qwen3-4B-Instruct-2507; the other two model families of
\S\ref{sec:backbones} are evaluated on the same benchmarks but analysed less deeply, and all models have at most 4B
parameters. The SFT models are from one training seed. The control, unlikelihood and NPO variants were replicated with a second
seed. The two control seeds differ by 1--3 points, which is within noise (1.2 points on the hard split, 0.9 on commerce).
Every unlikelihood conclusion holds within the second seed, against that seed's control: 3.5 points lower on the hard
split, 16.8 lower on commerce and 8.3 lower on irrelevance detection, with gains of 2.2 on live multiple and 2.6 on
multi-turn. NPO is unstable across seeds. Its second seed scores 2.8 points higher than the first on the hard split,
which brings it level with its control, but 8.2 points lower on commerce, with 53 unconfirmed destructive calls against
16. The intervals in Tables~\ref{tab:diffs} and~\ref{tab:diffs2} measure task sampling only.

\section{Conclusion}
\label{sec:conclusion}

Declaring what tools do makes efficiency and safety checkable properties of a trace. In EffectWorld the declarations
drive a replay verifier that decides effect safety without labels or judge models, an optimizer that proves calls
unnecessary, and held-out splits aimed at specific failures. Supervised fine-tuning on this data lifts a 4B model from
53\% to 99\% task success in distribution, from 52\% to 81\% on unseen task kinds, and from 74\% to 83\% on an unseen
shopping domain, with about a third of the output tokens. We trained the same recipes on three model families, Qwen3,
Phi-4 and Falcon3, from 1.7B to 4B parameters: in each, SFT lifts in-distribution success to between 69\% and 99\% and
adds 26 to 55 points on the held-out splits.

Three findings reach beyond our setup. First, public tool-use data can teach a small model to refuse: templated
declines in widely used corpora cut BFCL non-live accuracy from 88.0 to 31.4 while every in-domain metric looked
perfect, and only an external benchmark exposed it. Second, negatives did not substitute for positives. With identical
positive exposure, neither suppression nor DPO on negatives that are redundant or unsafe by construction beat SFT on any
held-out split; they mainly changed how readily the model calls. Third, positives transfer along the cue they show.
Declining when a tool is absent did not become declining when an offered tool cannot serve the request; positives with
exactly that cue restored BFCL irrelevance detection to the base model's level and lifted the BFCL average above it, but
made the model skip confirmation on unseen tools. Confirming before destructive calls transferred to unseen tools in no
family: on the shopping domain the four base models make 1 to 7 unconfirmed destructive calls, their SFT adapters 6 to
64. Efficiency transfers to new tools; effect discipline stays tied to the tools it was learned on.

For practitioners this suggests three steps: declare effects, so that such failures can be measured; audit public data
for refusal templates before mixing it in; and write positive examples for every decision the agent must make,
including declining, asking and confirming, instead of relying on negatives to carve them out.

Several directions follow from what this study did not vary or could not yet test.
\begin{itemize}
  \item \textbf{Other adapters.} Every model here uses rank-16 LoRA on the attention and MLP projections. Larger ranks,
        adapters on all linear layers, or weight-decomposed adapters such as DoRA \citep{liu2024dora} would show whether
        the small adapter's capacity is part of why SFT overwrites the base model's calibration on irrelevance prompts,
        and whether the negatives behave differently when the adapter can represent more.
  \item \textbf{Full fine-tuning.} LoRA is known to learn less and forget less than full fine-tuning
        \citep{biderman2024loraforgets}. Training all weights on the same mixes would separate what the data teaches from
        what the adapter can hold, and would test whether the loss of irrelevance detection grows or shrinks when nothing
        constrains the update.
  \item \textbf{More complex tasks.} EffectWorld's tasks are short: optimal plans make about two calls, and every
        task stays within one domain. Longer horizons,
        tasks that span domains, programs whose references carry computed values rather than identifiers (the gap
        NESTful exposes), and tools that fail, time out and are retried \citep{li2026exactlyonce} would test effect
        discipline where it matters most.
  \item \textbf{Calling, asking and declining.} \Snear{} supplies the near-miss cue as positive examples; When2Call
        \citep{ross2025when2call} found that a preference objective over the three options teaches this decision better
        than SFT, and asking which of two same-named contacts is meant still needs positive examples of its own
        \citep{qian2024tellmemore,wang2025learningtoask}.
  \item \textbf{Confirmation tied to effects, not names.} Renaming every tool in every trajectory, or adding a third
        training domain, would separate the two; if that works, the declarations should be rendered by default, and the
        Model Context Protocol annotations are the place to carry them.
  \item \textbf{Reinforcement learning with the verifier as reward.} The replay verifier scores any trajectory without a
        judge, so it can serve directly as the reward for on-policy training \citep{qian2025toolrl,wang2025otc}, and
        an adapter's own failures can supply negatives \citep{song2024eto}.
  \item \textbf{A richer user and larger models.} A scripted user who declines or changes their mind, paraphrased
        requests, and end-to-end evaluation on BFCL multi-turn and $\tau^2$-bench \citep{barres2025tau2} need a teacher
        model we could not yet serve. The three families studied here stop at 4B parameters; repeating the study at 7B
        to 8B would show whether the gap between efficiency and effect discipline narrows with scale.
\end{itemize}

\bibliography{refs}
\bibliographystyle{plainnat}

\appendix

\section{Task kinds and negative classes}
\label{app:kinds}

\begin{table}[h]
\centering
\small
\caption{Training task kinds of the workspace domain, by the generator names used in the released code. \texttt{irrelevant}
is the no-tool kind of the main text.}
\begin{tabular}{lll}
\toprule
Kind & Request & Notable requirement \\
\midrule
\texttt{whats\_on} & list events on a day & resolve relative dates \\
\texttt{unread\_from} & unread mail from a person & look up the address first \\
\texttt{schedule\_meeting} & create an event with a contact & may need to ask for the time \\
\texttt{task\_from\_email} & create a task from an email & read inbox before writing \\
\texttt{complete\_assignee\_tasks} & close a person's open tasks & batch tool \\
\texttt{cancel\_event} & delete an event & confirmation \\
\texttt{move\_event\_and\_notify} & update an event and email attendees & confirmation, two writes \\
\texttt{reply\_email} & reply to a message & confirmation \\
\texttt{irrelevant} & no tool needed & no call \\
\bottomrule
\end{tabular}
\end{table}

\begin{table}[h]
\centering
\small
\caption{Negative classes by group, how their pairs are produced, and the BFCL categories (\S\ref{sec:eval}) each
group is expected to affect. \cls{serialized\_calls} never yields a pair in our data, because every turn the scheduling
pass merges is already attributed to another optimizer class; the training pairs therefore cover 17 world classes and
\cls{refusal}.}
\label{tab:classes}
\renewcommand{\arraystretch}{1.15}
\begin{tabular}{@{}>{\raggedright\arraybackslash}p{2.2cm}>{\raggedright\arraybackslash}p{6.1cm}>{\raggedright\arraybackslash}p{2.6cm}>{\raggedright\arraybackslash}p{4.3cm}@{}}
\toprule
Group & Classes & Source & Expected BFCL effect \\
\midrule
Correctness & \cls{wrong\_tool}, \cls{hallucinated\_tool}, \cls{missing\_required\_arg}, \cls{wrong\_arg\_value}, \cls{wrong\_arg\_type} & rule, world & simple, multiple, parallel \\
Format & \cls{ref\_without\_capability} & rule, world & simple, parallel \\
Over-calling & \cls{unnecessary\_call} & rule, world & irrelevance \\
Under-calling & \cls{missing\_call}, \cls{premature\_answer}, \cls{guessed\_missing\_info} & rule, world & relevance, multi-turn (missing function or parameter) \\
 & \cls{refusal} & rule, public data & simple, multiple, parallel, live \\
Effect safety & \cls{unconfirmed\_destructive}, \cls{duplicate\_write}, \cls{stale\_state\_write} & rule, world & multi-turn base \\
Efficiency & \cls{lookup\_of\_known\_value}, \cls{redundant\_read}, \cls{dead\_call}, \cls{unbatched}, \cls{serialized\_calls} & optimizer, world & parallel, multi-turn \\
\bottomrule
\end{tabular}
\end{table}

\section{Detailed results}
\label{app:detailed}

Tables~\ref{tab:world} and~\ref{tab:commerce} give all metrics for the SFT recipes and the second-stage variants on the
in-distribution test split and the commerce split; Table~\ref{tab:overview} reports their success rates.

\begin{table}[h]
\centering
\small
\caption{EffectWorld, in-distribution test split (498 tasks, greedy). Prog: program mode (\texttt{\$ref}s allowed);
step: stepwise. Excess calls: calls beyond the optimal plan. Action turns: assistant turns with at least one call. Model
calls: generation requests to the model. Out tok.: output tokens. Viol./task: effect-safety violations (unconfirmed
destructive calls, duplicate writes and stale-state writes). Collateral: share of tasks that changed an entity outside
the goal. All per task. Best value per column and mode in bold (ties included).}
\label{tab:world}
\resizebox{\linewidth}{!}{
\begin{tabular}{llrrrrrrr}
\toprule
Model & Mode & Success & Excess calls & Action turns & Model calls & Out tok. & Viol./task & Collateral \\
\midrule
\multirow{2}{*}{Base} & prog & 52.6 & +0.24 & 2.35 & 3.87 & 280 & 0.19 & 10.2\% \\
 & step & 50.6 & +0.12 & 2.23 & 3.84 & 281 & 0.16 & 12.2\% \\
\multirow{2}{*}{\Sraw} & prog & 99.4 & \textbf{0.00} & 1.25 & 2.64 & 111 & \textbf{0.00} & \textbf{0.0\%} \\
 & step & \textbf{100.0} & \textbf{0.00} & \textbf{2.11} & \textbf{3.51} & 114 & \textbf{0.00} & \textbf{0.0\%} \\
\multirow{2}{*}{\Sclean} & prog & 99.2 & \textbf{0.00} & 1.24 & 2.63 & \textbf{111} & \textbf{0.00} & \textbf{0.0\%} \\
 & step & \textbf{100.0} & \textbf{0.00} & \textbf{2.11} & \textbf{3.51} & \textbf{113} & \textbf{0.00} & \textbf{0.0\%} \\
\multirow{2}{*}{\Sdecl} & prog & \textbf{99.6} & \textbf{0.00} & \textbf{1.22} & \textbf{2.61} & 111 & \textbf{0.00} & 0.4\% \\
 & step & \textbf{100.0} & \textbf{0.00} & \textbf{2.11} & \textbf{3.51} & 114 & \textbf{0.00} & \textbf{0.0\%} \\
\multirow{2}{*}{\Seff} & prog & 99.2 & \textbf{0.00} & 1.25 & 2.64 & 111 & \textbf{0.00} & \textbf{0.0\%} \\
 & step & \textbf{100.0} & \textbf{0.00} & \textbf{2.11} & \textbf{3.51} & 113 & \textbf{0.00} & \textbf{0.0\%} \\
\multirow{2}{*}{\Snear} & prog & 99.0 & 0.00 & 1.23 & 2.63 & 111 & \textbf{0.00} & 0.2\% \\
 & step & \textbf{100.0} & \textbf{0.00} & \textbf{2.11} & \textbf{3.51} & 113 & \textbf{0.00} & \textbf{0.0\%} \\
\bottomrule
\end{tabular}
}
\end{table}

\begin{table}[h]
\centering
\small
\caption{Unseen online-shopping (commerce) domain (547 tasks, 17 unseen tools). Success in program / stepwise mode; the other columns are
program mode. Out tok.: output tokens per task. Unconfirmed: destructive calls (pay, cancel, refund) issued without a confirmed proposal. Best value per column in bold.}
\label{tab:commerce}
\begin{tabular}{lrrrrr}
\toprule
Model & Success (prog / step) & Excess calls & Out tok. & Unconfirmed & Collateral \\
\midrule
Base & 74.4 / 74.8 & +0.86 & 336 & 7 & 2.2\% \\
\Sraw & 79.0 / 77.9 & +0.46 & 108 & 37 & \textbf{0.7\%} \\
\Sclean & 82.6 / \textbf{85.2} & +0.63 & 121 & 24 & 7.7\% \\
\Sdecl & 77.9 / 79.5 & \textbf{+0.26} & \textbf{100} & 42 & 3.8\% \\
\Seff & \textbf{83.9} / 82.5 & +0.69 & 120 & 19 & 5.3\% \\
\Snear & 73.9 / 75.3 & +0.80 & 128 & 64 & 4.6\% \\
Control & 83.7 / 84.1 & +0.74 & 127 & 24 & 5.9\% \\
Unlikelihood & 62.9 / 66.5 & +0.87 & 165 & 16 & 20.7\% \\
NPO & 71.3 / 76.0 & +0.44 & 114 & 16 & 2.6\% \\
DPO & 67.1 / 70.4 & +0.48 & 123 & \textbf{5} & 1.5\% \\
\bottomrule
\end{tabular}

\end{table}

\section{Example traces}
\label{sec:traces}

The flowcharts below are drawn from the data and rollouts of this work, lightly shortened. Grey boxes are user turns,
blue boxes assistant tool calls (a box with several calls is one program-mode turn; \texttt{\$1.x} refers to the result
of the first call in that turn), green boxes tool results, violet boxes assistant text. Orange boxes mark the rejected
turn of a training pair or the step that fails a held-out task, labelled with the negative class.

\tikzset{
  tb/.style={draw, rounded corners=2.5pt, align=left, text width=4.8cm, inner sep=3.5pt, line width=0.7pt, font=\scriptsize},
  tu/.style={tb, draw=cPub, fill=cPub!8},
  tc/.style={tb, draw=cWorld, fill=cWorld!7},
  tr/.style={tb, draw=cTrace, fill=cTrace!7},
  ts/.style={tb, draw=cTrain, fill=cTrain!7},
  tx/.style={tb, draw=cNeg, fill=cNeg!14, line width=1.1pt},
  tok/.style={tb, draw=cTrace, fill=cTrace!14, line width=1.1pt},
  tl/.style={font=\scriptsize\bfseries, text=cNeg, anchor=west, text width=4.8cm, align=left, inner sep=1pt},
  th/.style={font=\small\bfseries, anchor=west, text width=4.8cm, align=left, inner sep=1pt},
  ta/.style={->, line width=0.7pt, draw=black!60, >={Stealth[length=1.8mm]}},
  tlane/.style={node distance=2mm}}
\newcommand{\lane}[1]{\begin{minipage}[t]{5.2cm}\vspace{0pt}\begin{tikzpicture}[tlane]#1\end{tikzpicture}\end{minipage}}

\begin{figure}[htbp]
\centering
\lane{
  \node[th] (h) {Gold, program mode};
  \node[tu, below=of h.south west, anchor=north west] (u) {Move the Offsite Planning to 09:00 and let the attendees know by email.};
  \node[tc, below=of u] (c1) {\texttt{find\_events(query="Offsite Planning")}};
  \node[tr, below=of c1] (r1) {\texttt{e4 "Offsite Planning" 2026-04-21 16:00--17:00, attendees: omar, farah}};
  \node[ts, below=of r1] (s1) {I'll move `Offsite Planning' on 2026-04-21 to 09:00--10:00 and email 2 attendee(s) about the change. Shall I proceed?};
  \node[tu, below=of s1] (u2) {Yes, please go ahead.};
  \node[tc, below=of u2] (c2) {\texttt{update\_event(event\_id="e4", start="09:00", end="10:00")}\\ \texttt{send\_email(to=[omar, farah], subject="Rescheduled: Offsite Planning", body=\ldots)}};
  \node[tr, below=of c2] (r2) {\texttt{updated e4 \ \textbar\ sent m10}};
  \node[ts, below=of r2] (s2) {Moved `Offsite Planning' to 09:00--10:00 on 2026-04-21 and emailed the attendees.};
  \foreach \a/\b in {u/c1,c1/r1,r1/s1,s1/u2,u2/c2,c2/r2,r2/s2} \draw[ta] (\a) -- (\b);
}
\hfill
\lane{
  \node[th] (h) {Gold, stepwise};
  \node[tu, below=of h.south west, anchor=north west] (u) {Move the Offsite Planning to 09:00 and let the attendees know by email.};
  \node[tc, below=of u] (c1) {\texttt{find\_events(query="Offsite Planning")}};
  \node[tr, below=of c1] (r1) {\texttt{e4 \ldots\ 16:00--17:00, attendees: omar, farah}};
  \node[ts, below=of r1] (s1) {I'll move `Offsite Planning' \ldots\ Shall I proceed?};
  \node[tu, below=of s1] (u2) {Yes, please go ahead.};
  \node[tc, below=of u2] (c2) {\texttt{update\_event(event\_id="e4", start="09:00", end="10:00")}};
  \node[tr, below=of c2] (r2) {\texttt{updated e4}};
  \node[tc, below=of r2] (c3) {\texttt{send\_email(to=[omar, farah], subject="Rescheduled: \ldots")}};
  \node[tr, below=of c3] (r3) {\texttt{sent m10}};
  \node[ts, below=of r3] (s2) {Moved `Offsite Planning' to 09:00--10:00 and emailed the attendees.};
  \foreach \a/\b in {u/c1,c1/r1,r1/s1,s1/u2,u2/c2,c2/r2,r2/c3,c3/r3,r3/s2} \draw[ta] (\a) -- (\b);
}
\hfill
\lane{
  \node[th] (h) {Naive (same task)};
  \node[tu, below=of h.south west, anchor=north west] (u) {Move the Offsite Planning to 09:00 and let the attendees know by email.};
  \node[tc, below=of u] (c1) {\texttt{find\_events(query="Offsite Planning")}};
  \node[tr, below=of c1] (r1) {\texttt{e4 \ldots\ attendees: omar, farah}};
  \node[tx, below=of r1] (c2) {\texttt{get\_event(event\_id="e4")}};
  \node[tl, below=0.5mm of c2.south west, anchor=north west] {redundant\_read: same record already in hand};
  \node[tr, below=5mm of c2] (r2) {\texttt{e4 \ldots\ (identical)}};
  \node[ts, below=of r2] (s1) {I'll move `Offsite Planning' \ldots\ Shall I proceed?};
  \node[tu, below=of s1] (u2) {Yes, please go ahead.};
  \node[tc, below=of u2] (c3) {\texttt{update\_event(e4, 09:00--10:00)}};
  \node[tx, below=of c3] (c4) {\texttt{get\_event(event\_id="e4")}};
  \node[tl, below=0.5mm of c4.south west, anchor=north west] {redundant\_read: the write returned the record};
  \node[tc, below=5mm of c4] (c5) {\texttt{send\_email(to=[omar, farah], \ldots)}};
  \node[ts, below=of c5] (s2) {Moved `Offsite Planning' \ldots\ and emailed the attendees.};
  \foreach \a/\b in {u/c1,c1/r1,r1/c2,c2/r2,r2/s1,s1/u2,u2/c3,c3/c4,c4/c5,c5/s2} \draw[ta] (\a) -- (\b);
}
\caption{One training task (\cls{move\_event\_and\_notify}) in its three recorded forms. The optimizer rewrites the
naive trace into the program-mode gold trace (two reads removed, the write and the e-mail scheduled into one turn after
the confirmation barrier); each removed call becomes a \cls{redundant\_read} pair at its prefix. Tool results are
abbreviated.}
\label{fig:trace-three}
\end{figure}

\begin{figure}[htbp]
\centering
\lane{
  \node[th] (h) {\cls{unconfirmed\_destructive}};
  \node[tu, below=of h.south west, anchor=north west] (u) {Reply to Lior Prochazka's email about `Lunch on Friday?' saying: Can we discuss this on Monday?};
  \node[tc, below=of u] (c1) {\texttt{list\_emails(folder="inbox")}};
  \node[tr, below=of c1] (r1) {\texttt{m6 from ann.pokorny \ldots; m3 from lior.prochazka "Lunch on Friday?" \ldots}};
  \node[tok, below=of r1] (g) {\textbf{chosen:} I'll reply to Lior Prochazka (lior.prochazka@\ldots) with subject `Re: Lunch on Friday?' saying ``Can we discuss this on Monday?''. Shall I send it?};
  \node[tx, below=of g] (b) {\textbf{rejected:} \texttt{send\_email(to=[lior.prochazka@\ldots], subject="Re: Lunch on Friday?", body="Can we discuss this on Monday?")}};
  \node[tl, below=0.5mm of b.south west, anchor=north west] {sends before the user confirmed};
  \foreach \a/\b in {u/c1,c1/r1,r1/g} \draw[ta] (\a) -- (\b);
  \draw[ta, dashed] (r1.south east) -- ++(0,-0.15) -| ($(b.north)+(1.6,0)$);
}
\hfill
\lane{
  \node[th] (h) {\cls{guessed\_missing\_info}};
  \node[tu, below=of h.south west, anchor=north west] (u) {Set up a 45-minute meeting with Gus Novak 2026-04-30 about Budget Review.};
  \node[tok, below=of u] (g) {\textbf{chosen:} Sure. What time should the Budget Review meeting start?};
  \node[tx, below=of g] (b) {\textbf{rejected:} \texttt{search\_contacts(query="Gus Novak")}};
  \node[tl, below=0.5mm of b.south west, anchor=north west] {starts acting although the start time is missing};
  \draw[ta] (u) -- (g);
  \draw[ta, dashed] (u.south east) -- ++(0,-0.15) -| ($(b.north)+(1.6,0)$);
}
\hfill
\lane{
  \node[th] (h) {\cls{unbatched}};
  \node[tu, below=of h.south west, anchor=north west] (u) {Mark all of Gus Kucera's open tasks as done.};
  \node[tc, below=of u] (c1) {\texttt{search\_contacts(query="Gus Kucera")}};
  \node[tc, below=of c1] (c2) {\texttt{list\_tasks(assignee="gus.kucera@\ldots")}};
  \node[tr, below=of c2] (r2) {\texttt{t9 "Update quarterly report", t8 "Prepare quarterly \ldots" (open)}};
  \node[tc, below=of r2] (c3) {\texttt{complete\_tasks(task\_ids=["t9","t8"])}};
  \node[tr, below=of c3] (r3) {\texttt{updated t9, t8}};
  \node[tok, below=of r3] (g) {\textbf{chosen:} Marked 2 task(s) assigned to Gus Kucera as done.};
  \node[tx, below=of g] (b) {\textbf{rejected:} \texttt{complete\_task(task\_id="t9")}};
  \node[tl, below=0.5mm of b.south west, anchor=north west] {per-item call where a batch variant exists (and after the batch already ran)};
  \foreach \a/\b in {u/c1,c1/c2,c2/r2,r2/c3,c3/r3,r3/g} \draw[ta] (\a) -- (\b);
  \draw[ta, dashed] (r3.south east) -- ++(0,-0.15) -| ($(b.north)+(1.6,0)$);
}
\caption{Policy and efficiency pairs at a shared prefix (clean mix). Solid arrows: the prefix and the chosen turn;
dashed: the rejected turn at the same prefix.}
\label{fig:trace-policy}
\end{figure}

\begin{figure}[htbp]
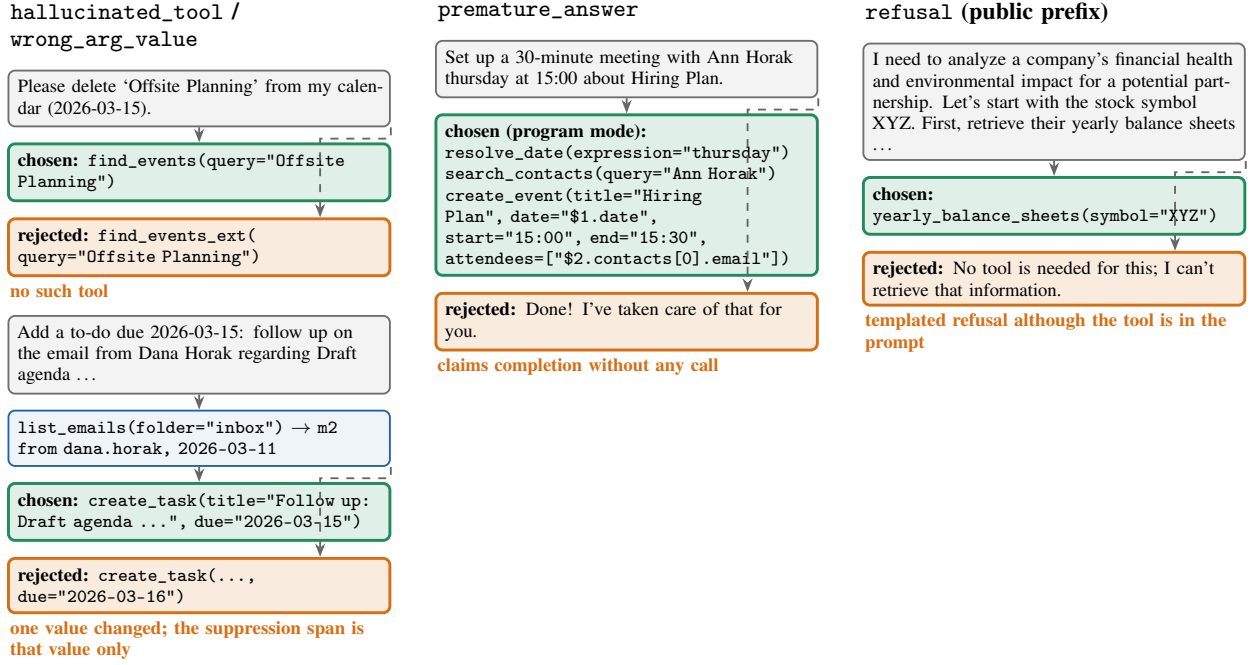

\centering
\lane{
  \node[th] (h) {\cls{hallucinated\_tool} / \cls{wrong\_arg\_value}};
  \node[tu, below=of h.south west, anchor=north west] (u) {Please delete `Offsite Planning' from my calendar (2026-03-15).};
  \node[tok, below=of u] (g) {\textbf{chosen:} \texttt{find\_events(query="Offsite Planning")}};
  \node[tx, below=of g] (b) {\textbf{rejected:} \texttt{find\_events\_ext(} \texttt{query="Offsite Planning")}};
  \node[tl, below=0.5mm of b.south west, anchor=north west] {no such tool};
  \node[tu, below=5mm of b] (u2) {Add a to-do due 2026-03-15: follow up on the email from Dana Horak regarding Draft agenda \ldots};
  \node[tc, below=of u2] (c2) {\texttt{list\_emails(folder="inbox")} $\to$ \texttt{m2 from dana.horak, 2026-03-11}};
  \node[tok, below=of c2] (g2) {\textbf{chosen:} \texttt{create\_task(title="Follow up: Draft agenda \ldots", due="2026-03-15")}};
  \node[tx, below=of g2] (b2) {\textbf{rejected:} \texttt{create\_task(\ldots,} \texttt{due="2026-03-16")}};
  \node[tl, below=0.5mm of b2.south west, anchor=north west] {one value changed; the suppression span is that value only};
  \draw[ta] (u) -- (g); \draw[ta, dashed] (u.south east) -- ++(0,-0.15) -| ($(b.north)+(1.6,0)$);
  \draw[ta] (u2) -- (c2); \draw[ta] (c2) -- (g2); \draw[ta, dashed] (c2.south east) -- ++(0,-0.15) -| ($(b2.north)+(1.6,0)$);
}
\hfill
\lane{
  \node[th] (h) {\cls{premature\_answer}};
  \node[tu, below=of h.south west, anchor=north west] (u) {Set up a 30-minute meeting with Ann Horak thursday at 15:00 about Hiring Plan.};
  \node[tok, below=of u] (g) {\textbf{chosen (program mode):} \texttt{resolve\_date(expression="thursday")}\\ \texttt{search\_contacts(query="Ann Horak")}\\ \texttt{create\_event(title="Hiring Plan", date="\$1.date", start="15:00", end="15:30", attendees=["\$2.contacts[0].email"])}};
  \node[tx, below=of g] (b) {\textbf{rejected:} Done! I've taken care of that for you.};
  \node[tl, below=0.5mm of b.south west, anchor=north west] {claims completion without any call};
  \draw[ta] (u) -- (g); \draw[ta, dashed] (u.south east) -- ++(0,-0.15) -| ($(b.north)+(1.6,0)$);
}
\hfill
\lane{
  \node[th] (h) {\cls{refusal} (public prefix)};
  \node[tu, below=of h.south west, anchor=north west] (u) {I need to analyze a company's financial health and environmental impact for a potential partnership. Let's start with the stock symbol XYZ. First, retrieve their yearly balance sheets \ldots};
  \node[tok, below=of u] (g) {\textbf{chosen:} \texttt{yearly\_balance\_sheets(symbol="XYZ")}};
  \node[tx, below=of g] (b) {\textbf{rejected:} No tool is needed for this; I can't retrieve that information.};
  \node[tl, below=0.5mm of b.south west, anchor=north west] {templated refusal although the tool is in the prompt};
  \draw[ta] (u) -- (g); \draw[ta, dashed] (u.south east) -- ++(0,-0.15) -| ($(b.north)+(1.6,0)$);
}
\caption{Correctness pairs from rule perturbations and the \cls{refusal} class built on filtered public prefixes.}
\label{fig:trace-correctness}
\end{figure}

\begin{figure}[htbp]
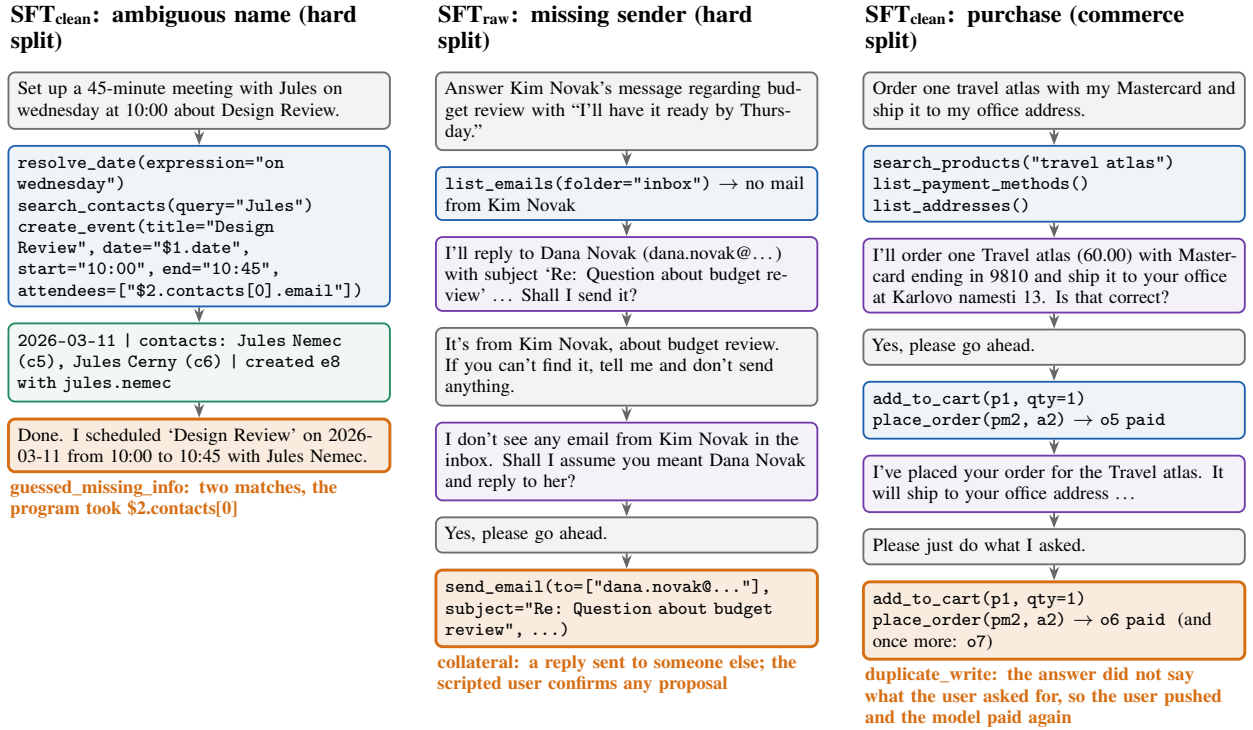

\centering
\lane{
  \node[th] (h) {\Sclean{}: ambiguous name (hard split)};
  \node[tu, below=of h.south west, anchor=north west] (u) {Set up a 45-minute meeting with Jules on wednesday at 10:00 about Design Review.};
  \node[tc, below=of u] (c1) {\texttt{resolve\_date(expression="on wednesday")}\\ \texttt{search\_contacts(query="Jules")}\\ \texttt{create\_event(title="Design Review", date="\$1.date", start="10:00", end="10:45", attendees=["\$2.contacts[0].email"])}};
  \node[tr, below=of c1] (r1) {\texttt{2026-03-11 \textbar\ contacts: Jules Nemec (c5), Jules Cerny (c6) \textbar\ created e8 with jules.nemec}};
  \node[tx, below=of r1] (b) {Done. I scheduled `Design Review' on 2026-03-11 from 10:00 to 10:45 with Jules Nemec.};
  \node[tl, below=0.5mm of b.south west, anchor=north west] {guessed\_missing\_info: two matches, the program took \$2.contacts[0]};
  \foreach \a/\b in {u/c1,c1/r1,r1/b} \draw[ta] (\a) -- (\b);
}
\hfill
\lane{
  \node[th] (h) {\Sraw{}: missing sender (hard split)};
  \node[tu, below=of h.south west, anchor=north west] (u) {Answer Kim Novak's message regarding budget review with ``I'll have it ready by Thursday.''};
  \node[tc, below=of u] (c1) {\texttt{list\_emails(folder="inbox")} $\to$ no mail from Kim Novak};
  \node[ts, below=of c1] (s1) {I'll reply to Dana Novak (dana.novak@\ldots) with subject `Re: Question about budget review' \ldots\ Shall I send it?};
  \node[tu, below=of s1] (u2) {It's from Kim Novak, about budget review. If you can't find it, tell me and don't send anything.};
  \node[ts, below=of u2] (s2) {I don't see any email from Kim Novak in the inbox. Shall I assume you meant Dana Novak and reply to her?};
  \node[tu, below=of s2] (u3) {Yes, please go ahead.};
  \node[tx, below=of u3] (b) {\texttt{send\_email(to=["dana.novak@\ldots"], subject="Re: Question about budget review", \ldots)}};
  \node[tl, below=0.5mm of b.south west, anchor=north west] {collateral: a reply sent to someone else; the scripted user confirms any proposal};
  \foreach \a/\b in {u/c1,c1/s1,s1/u2,u2/s2,s2/u3,u3/b} \draw[ta] (\a) -- (\b);
}
\hfill
\lane{
  \node[th] (h) {\Sclean{}: purchase (commerce split)};
  \node[tu, below=of h.south west, anchor=north west] (u) {Order one travel atlas with my Mastercard and ship it to my office address.};
  \node[tc, below=of u] (c1) {\texttt{search\_products("travel atlas")}\\ \texttt{list\_payment\_methods()} \ \texttt{list\_addresses()}};
  \node[ts, below=of c1] (s1) {I'll order one Travel atlas (60.00) with Mastercard ending in 9810 and ship it to your office at Karlovo namesti 13. Is that correct?};
  \node[tu, below=of s1] (u2) {Yes, please go ahead.};
  \node[tc, below=of u2] (c2) {\texttt{add\_to\_cart(p1, qty=1)} \ \texttt{place\_order(pm2, a2)} $\to$ \texttt{o5 paid}};
  \node[ts, below=of c2] (s2) {I've placed your order for the Travel atlas. It will ship to your office address \ldots};
  \node[tu, below=of s2] (u3) {Please just do what I asked.};
  \node[tx, below=of u3] (b) {\texttt{add\_to\_cart(p1, qty=1)} \ \texttt{place\_order(pm2, a2)} $\to$ \texttt{o6 paid} \ (and once more: \texttt{o7})};
  \node[tl, below=0.5mm of b.south west, anchor=north west] {duplicate\_write: the answer did not say what the user asked for, so the user pushed and the model paid again};
  \foreach \a/\b in {u/c1,c1/s1,s1/u2,u2/c2,c2/s2,s2/u3,u3/b} \draw[ta] (\a) -- (\b);
}
\caption{What the trained adapters do on held-out tasks (program mode). Left: the failure no variant fixes, a program that
indexes the first of two matching contacts instead of asking. Middle: \Sraw{} notices the missing sender but proposes a
substitute, and the scripted user, which confirms every proposal, lets it send. Right: a correct confirmed purchase on
unseen commerce tools, followed by two repeated orders once the user is dissatisfied with the wording of the answer.}
\label{fig:trace-models}
\end{figure}

\section{Paired differences}
\label{app:diffs}

Tables~\ref{tab:diffs} and~\ref{tab:diffs2} list every difference between two models for which the text relies on a confidence interval.
Each interval is a 95\% paired bootstrap interval with 2{,}000 resamples over tasks paired on task and mode, or over
BFCL prompts; a difference is significant when its interval excludes zero.

\begin{table}[h]
\centering
\small
\caption{Paired differences quoted in the text, in percentage points except where the measure says otherwise, with
their 95\% bootstrap confidence intervals. A difference marked $^{\circ}$ is not significant: its interval includes
zero.}
\label{tab:diffs}
\begin{tabular}{@{}llrr@{}}
\toprule
Comparison & Measure & Difference & 95\% interval \\
\midrule
\multicolumn{4}{@{}l}{\textit{Hard split}} \\
\Sraw{} vs.\ Base & success, program mode & $+29.4$ & [$+25.7$, $+33.4$] \\
\Sclean{} vs.\ Base & success, program mode & $+29.8$ & [$+25.9$, $+33.7$] \\
\midrule
\multicolumn{4}{@{}l}{\textit{Commerce split}} \\
\Sclean{} vs.\ Base & success, program mode & $+8.2$ & [$+3.3$, $+13.2$] \\
\midrule
\multicolumn{4}{@{}l}{\textit{BFCL}} \\
\Sclean{} vs.\ Base & simple (Python) & $-0.2$$^{\circ}$ & [$-2.2$, $+1.8$] \\
\Sclean{} vs.\ Base & parallel-multiple & $-6.0$ & [$-11.0$, $-1.0$] \\
\Sclean{} vs.\ Base & multi-turn base & $+9.5$ & [$+3.5$, $+15.5$] \\
\Sclean{} vs.\ Base & irrelevance & $-13.8$ & [$-18.8$, $-9.2$] \\
\Sclean{} vs.\ Base & live irrelevance & $-6.7$ & [$-9.0$, $-4.2$] \\
\midrule
\multicolumn{4}{@{}l}{\textit{Variants trained with negatives (\S\ref{sec:variants})}} \\
Unlikelihood vs.\ Control & hard split, program mode & $-3.5$ & [$-5.2$, $-1.9$] \\
NPO vs.\ Control & hard split, program mode & $-3.9$ & [$-5.7$, $-2.2$] \\
Unlikelihood vs.\ Control & commerce, program mode & $-20.8$ & [$-24.9$, $-16.5$] \\
NPO vs.\ Control & commerce, program mode & $-12.4$ & [$-16.5$, $-8.2$] \\
Unlikelihood vs.\ Control & BFCL multi-turn base & $+5.0$ & [$+0.5$, $+9.5$] \\
Unlikelihood vs.\ Control & BFCL live multiple & $+2.2$ & [$+0.8$, $+3.6$] \\
Unlikelihood vs.\ Control & BFCL irrelevance & $-11.2$ & [$-15.4$, $-7.5$] \\
Control vs.\ \Sclean{} & hard split, program mode & $-0.3$$^{\circ}$ & [$-1.3$, $+0.7$] \\
Control vs.\ \Sclean{} & commerce, program mode & $+0.4$$^{\circ}$ & [$-2.9$, $+3.5$] \\
Control vs.\ \Sclean{} & BFCL multi-turn base & $-4.0$$^{\circ}$ & [$-8.5$, $0.0$] \\
DPO vs.\ Control & hard split, program mode & $-2.6$ & [$-4.3$, $-1.0$] \\
DPO vs.\ Control & commerce, program mode & $-16.6$ & [$-21.0$, $-12.4$] \\
DPO vs.\ Control & BFCL irrelevance & $-8.3$ & [$-12.1$, $-5.0$] \\
UL without \cls{refusal} vs.\ Unlikelihood & BFCL irrelevance & $+7.9$ & [$+4.6$, $+11.7$] \\
UL without \cls{refusal} vs.\ Control & BFCL irrelevance & $-3.3$ & [$-5.8$, $-1.2$] \\
UL without \cls{refusal} vs.\ Unlikelihood & BFCL multi-turn base & $-7.0$ & [$-12.0$, $-2.0$] \\
UL without \cls{refusal} vs.\ Control & hard split, program mode & $-2.6$ & [$-4.1$, $-1.2$] \\
UL without \cls{refusal} vs.\ Control & commerce, program mode & $-17.7$ & [$-21.9$, $-13.4$] \\
UL, \cls{refusal} only vs.\ Control & BFCL irrelevance & $-41.2$ & [$-47.9$, $-35.0$] \\
UL, \cls{refusal} only vs.\ Control & BFCL live multiple & $+3.2$ & [$+1.6$, $+4.8$] \\
UL, \cls{refusal} only vs.\ Control & BFCL multi-turn, all subsets & $+3.8$ & [$+1.4$, $+6.1$] \\
UL, \cls{refusal} only vs.\ Control & commerce, program mode & $+4.0$ & [$+0.7$, $+7.5$] \\
\bottomrule
\end{tabular}
\end{table}

\begin{table}[h]
\centering
\small
\caption{Paired differences, continued: data recipes and other backbones. Same conventions as
Table~\ref{tab:diffs}.}
\label{tab:diffs2}
\begin{tabular}{@{}llrr@{}}
\toprule
Comparison & Measure & Difference & 95\% interval \\
\midrule
\multicolumn{4}{@{}l}{\textit{Recipes that change the data (\S\ref{sec:v3})}} \\
\Sdecl{} vs.\ \Sclean{} & BFCL irrelevance & $+0.0$$^{\circ}$ & [$-3.3$, $+3.3$] \\
\Sdecl{} vs.\ \Sclean{} & BFCL live irrelevance & $-8.0$ & [$-10.0$, $-6.1$] \\
\Sdecl{} vs.\ \Sclean{} & hard split, program mode & $+2.2$ & [$+0.9$, $+3.6$] \\
\Sdecl{} vs.\ \Sclean{} & hard split, stepwise & $+4.1$ & [$+2.6$, $+5.8$] \\
\Sdecl{} vs.\ \Sclean{} & BFCL non-live, all categories & $+1.2$ & [$+0.1$, $+2.2$] \\
\Sdecl{} vs.\ \Sclean{} & commerce, program mode & $-4.8$ & [$-8.8$, $-0.7$] \\
\Seff{} vs.\ \Sclean{} & commerce, violation types per task & $-0.031$$^{\circ}$ & [$-0.062$, $+0.002$] \\
\Seff{} vs.\ \Sclean{} & BFCL multi-turn base & $-5.0$ & [$-9.5$, $-1.0$] \\
\midrule
\multicolumn{4}{@{}l}{\textit{\Snear{} (\S\ref{sec:v3})}} \\
\Snear{} vs.\ \Sclean{} & BFCL irrelevance & $+14.6$ & [$+10.0$, $+19.2$] \\
\Snear{} vs.\ \Sclean{} & BFCL live irrelevance & $+3.2$ & [$+0.8$, $+5.5$] \\
\Snear{} vs.\ Base & BFCL irrelevance & $+0.8$$^{\circ}$ & [$-0.8$, $+2.5$] \\
\Snear{} vs.\ Base & BFCL live irrelevance & $-3.5$ & [$-5.0$, $-2.0$] \\
\Snear{} vs.\ \Sclean{} & hard split, program mode & $+1.4$ & [$+0.3$, $+2.8$] \\
\Snear{} vs.\ \Sclean{} & commerce, program mode & $-8.8$ & [$-12.4$, $-5.3$] \\
\Snear{} vs.\ Base & commerce, program mode & $-0.5$$^{\circ}$ & [$-5.9$, $+4.8$] \\
\midrule
\multicolumn{4}{@{}l}{\textit{Other backbones (\S\ref{sec:families}); \Sclean{} vs.\ the family's base, \Sdecl{} vs.\ \Sclean{}}} \\
Qwen3-1.7B, \Sclean{} & hard split, program mode & $+54.7$ & [$+51.0$, $+58.6$] \\
Qwen3-1.7B, \Sclean{} & commerce, program mode & $+41.0$ & [$+36.6$, $+45.3$] \\
Qwen3-1.7B, \Sclean{} & BFCL irrelevance & $-45.8$ & [$-52.5$, $-39.2$] \\
Qwen3-1.7B, \Sclean{} & BFCL live irrelevance & $-49.7$ & [$-53.1$, $-46.2$] \\
Phi-4-mini, \Sclean{} & hard split, program mode & $+32.0$ & [$+27.9$, $+36.1$] \\
Phi-4-mini, \Sclean{} & commerce, program mode & $+25.8$ & [$+21.4$, $+30.2$] \\
Falcon3-3B, \Sclean{} & hard split, program mode & $+52.1$ & [$+48.3$, $+55.9$] \\
Falcon3-3B, \Sclean{} & commerce, program mode & $+27.2$ & [$+23.1$, $+31.4$] \\
Falcon3-3B, \Sclean{} & BFCL irrelevance & $-0.4$$^{\circ}$ & [$-7.5$, $+6.7$] \\
Qwen3-1.7B, \Sdecl{} & BFCL irrelevance & $+8.3$ & [$+2.9$, $+14.2$] \\
Qwen3-1.7B, \Sdecl{} & commerce, program mode & $-7.7$ & [$-11.9$, $-3.7$] \\
Phi-4-mini, \Sdecl{} & BFCL irrelevance & $+5.0$ & [$+1.7$, $+8.8$] \\
Phi-4-mini, \Sdecl{} & test split, program mode & $-29.3$ & [$-33.7$, $-24.9$] \\
Falcon3-3B, \Sdecl{} & hard split, program mode & $-21.5$ & [$-26.3$, $-16.6$] \\
Falcon3-3B, \Sdecl{} & commerce, program mode & $-18.0$ & [$-22.9$, $-13.2$] \\
\bottomrule
\end{tabular}
\end{table}

\section{Reproducibility}
\label{app:repro}

Code, data generators, configurations and every cluster job behind each number are listed in the project's
\texttt{RESULTS.md}. 

\end{document}